\documentclass[a4paper,11pt]{article}
\pdfoutput=1 

\usepackage{jheppub} 

\usepackage[T1]{fontenc} 
\usepackage{graphicx,color}
\usepackage{amsmath,amssymb}
\usepackage{url}
\usepackage{epstopdf}
\usepackage[utf8]{inputenc}

\newcommand{\be}{\begin{equation}}
\newcommand{\ee}{\end{equation}}
\newcommand{\bea}{\begin{eqnarray}}
\newcommand{\eea}{\end{eqnarray}}

\newcommand{\bc}{\begin{center}}
	\newcommand{\ec}{\end{center}}

\newcommand {\ba}{\begin{array}}
	\newcommand {\ea}{\end{array}}
\newcommand{\ben}{\begin{enumerate}}
	\newcommand{\een}{\end{enumerate}}

\usepackage{epsfig,graphicx}
\usepackage{bm}
\usepackage{dcolumn}

\newcommand{\Antonio}[1]{{\color{blue}#1}}

\begin{document}

\title{\boldmath A variant of 3-3-1 model for the generation of the SM fermion mass and mixing pattern}

\author[a]{A. E. C\'{a}rcamo Hern\'{a}ndez,}
\author[a]{Sergey Kovalenko,}
\author[b,c,1]{H. N. Long,}

\author[a]{Ivan Schmidt}

\affiliation[a]{Universidad T\'{e}cnica Federico Santa Mar\'{\i}a\\
and Centro Cient\'{\i}fico-Tecnol\'{o}gico de Valpara\'{\i}so\\
Casilla 110-V, Valpara\'{\i}so, Chile}
\affiliation[b]{Theoretical Particle Physics and Cosmology Research Group, Advanced Institute for Materials Science, Ton Duc Thang University, Ho Chi Minh City, Vietnam}
\affiliation[c]{Faculty of Applied Sciences, Ton Duc Thang University, Ho Chi Minh City, Vietnam}

\emailAdd{antonio.carcamo@usm.cl}
\emailAdd{sergey.kovalenko@usm.cl}
\emailAdd{hoangngoclong@tdt.edu.vn}
\emailAdd{ivan.schmidt@usm.cl}

\abstract{We propose an extension of the 3-3-1 model with an additional symmetry group
\mbox{$Z_{2}\times Z_{4} \times U(1)_{L_g}$}
and an extended scalar sector.
To our best knowledge this is the first example of a renormalizable 3-3-1 model,
which allows explanation of the SM fermion mass hierarchy by a sequential loop
suppression: tree-level top and exotic fermion masses, 1-loop bottom, charm,
tau and muon masses; 2-loop masses for the light up, down, strange quarks as
well as for the electron. The light active neutrino masses are
generated from a combination of linear and inverse seesaw mechanisms at two
loop level. The model also has viable fermionic and scalar dark matter candidates.}

\arxivnumber{1705.09169}

\newpage

\maketitle
\flushbottom

\allowdisplaybreaks


\section{Introduction \label{intro}}

Despite the great consistency of the Standard Model (SM) with
experimental data, recently confirmed by the LHC discovery of the $126$ GeV
Higgs boson \cite{Aad:2012tfa,Chatrchyan:2012xdj}, it has several
unexplained issues \cite{Agashe:2014kda}. Among the most pressings ones are the
smallness of neutrino masses, the fermion mass and mixing hierarchy, and the
existence of three fermion families.

In the SM the flavor structure of the Yukawa interactions is not restricted
by gauge invariance. Consequently, fermion masses and mixings are left
unfixed, and the SM does not provide an explanation for their large hierarchy,
which spreads over a range of five orders of magnitude in
the quark sector, and a dramatically broader range of about 11 orders of
magnitude, if we include the neutrinos. Even though in the SM these
parameters appear only through Yukawa interaction terms and not in explicit
mass terms, this mechanism does not provide an explanation for their values,
but only translates the problem to fitting different Yukawa couplings, one
for each mass and with disparate values for some of them. The origin of
quark mixing and the size of CP violation in this sector is also a related
issue. A fundamental theory is expected to provide a dynamical explanation
for the masses and mixings.

While the mixing angles in the quark sector are very small, in the lepton
sector two of the mixing angles are large, and one mixing angle is small.
This suggests a different kind of New Physics for the neutrino sector from
the one present in the quark mass and mixing pattern. Experiments with
solar, atmospheric and reactor neutrinos have brought clear evidence of
neutrino oscillations from the measured non vanishing neutrino mass squared
splittings. This brings compelling and indubitable evidence that at least two
of the neutrinos have non vanishing masses, much smaller, by many orders of
magnitude, than the SM charged fermion masses, and that the three neutrino
flavors mix with each other.

The flavor puzzle of the SM indicates that New Physics has to be advocated in order
to explain the prevailing pattern of fermion masses and mixings. To tackle
the limitations of the SM, various extensions, including larger scalar
and/or fermion sectors, as well as extended gauge groups with additional flavor
symmetries, have been proposed in the literature \cite%
{Patel:2010hr,Morisi:2011pm,Gupta:2011ct,BhupalDev:2012nm,King:2013hj,Aranda:2013gga,Gomez-Izquierdo:2013uaa,Hernandez:2013dta,Campos:2014lla,Bonilla:2014xla,Campos:2014zaa,Arbelaez:2015toa,Hernandez:2015dga,Hernandez:2015zeh,Hernandez:2015hrt,Joshipura:2015dsa,Bjorkeroth:2015uou,Bjorkeroth:2015ora,Varzielas:2015aua,Chen:2015jta,Ma:2015fpa,Ma:2016nkf,Arbelaez:2016mhg,Pasquini:2016kwk,Carballo-Perez:2016ooy,Vien:2016qbb,Chulia:2016giq,Penedo:2017vtf,Cruz:2017add,Vien:2013zra,Hernandez:2013hea,Vien:2014pta,Vien:2014gza,Vien:2014vka, Vien:2014ica,Hernandez:2014lpa,Hernandez:2014vta,VanVien:2015xha,Hernandez:2015tna,Hernandez:2015cra,Vien:2016tmh,Hernandez:2016eod,Camargo-Molina:2016yqm,Camargo-Molina:2016bwm,Gomez-Izquierdo:2017rxi,Gomez-Izquierdo:2017med,Chattopadhyay:2017zvs,CarcamoHernandez:2017kra,CarcamoHernandez:2017owh,Bernal:2017xat,CarcamoHernandez:2018iel,deAnda:2018oik,CarcamoHernandez:2018aon}%
. Recent reviews on flavor symmetries are provided in Refs. \cite%
{Altarelli:2002hx,Altarelli:2010gt,Ishimori:2010au,King:2013eh,King:2014nza,King:2017guk}%
. 
Another approach to describe the fermion mass and mixing pattern consists in
postulating particular mass matrix textures (see Refs \cite%
{Fritzsch:1977za,Fukuyama:1997ky,Du:1992iy,Barbieri:1994kw,Peccei:1995fg,Fritzsch:1999ee,Roberts:2001zy,Nishiura:2002ei,deMedeirosVarzielas:2005ax,Carcamo:2006dp,Kajiyama:2007gx,CarcamoHernandez:2010im,Branco:2010tx,Leser:2011fz,Gupta:2012dma,Dev:2012sg,CarcamoHernandez:2012xy,Hernandez:2013mcf,Pas:2014bra,Hernandez:2014hka,Hernandez:2014zsa,Nishiura:2014psa,Frank:2014aca,Ghosal:2015lwa,Sinha:2015ooa,Nishiura:2015qia,Samanta:2015oqa,Gautam:2015kya,Pas:2015hca,Gupta:2013yha,Wang:2015saa}
for works which consider textures). In addition, the hierarchy of SM
charged fermion masses can also be explained by considering the charged
fermion Yukawa matrices as products of a few random matrices, which
typically feature strong hierarchies in their eigenvalue spectrum, even though
the individual entries are of order unity, as was recently observed in Ref. \cite%
{vonGersdorff:2017iym}.

Concerning models with an extended gauge symmetry, those based on the gauge
symmetry $SU(3)_{c}\times SU(3)_{L}\times U(1)_{X}$, also called 3-3-1
models, which introduce a family non-universal $U(1)_{X}$ symmetry \cite%
{Georgi:1978bv,Valle:1983dk,Pisano:1991ee,Foot:1992rh,Frampton:1992wt,Hoang:1996gi,Hoang:1995vq,Foot:1994ym,Boucenna:2015pav,Hernandez:2015ywg,Fonseca:2016tbn,Hati:2017aez}%
, can provide an explanation for the origin of the family structure of the
fermions. These models have the following phenomenological advantages: (%
\textbf{i}) The three family structure in the fermion sector can be
understood in the 3-3-1 models from the cancellation of chiral anomalies and
asymptotic freedom in QCD. (\textbf{ii}) The fact that the third family is
treated under a different representation can explain the large mass
difference between the heaviest quark family and the two lighter ones. (%
\textbf{iii}) The 3-3-1 models allow for the quantization of electric charge
\cite{deSousaPires:1998jc,VanDong:2005ux}. (\textbf{iv}) These models have
several sources of CP violation \cite{Montero:1998yw,Montero:2005yb}. (%
\textbf{v}) These models explain why the Weinberg mixing angle satisfies $%
\sin^2\theta_W<\frac1{4}$. (\textbf{vi}) These models contain a natural
Peccei-Quinn symmetry, which solves the strong-CP problem \cite%
{Pal:1994ba,Dias:2002gg,Dias:2003zt,Dias:2003iq}. (\textbf{vii}) The 3-3-1
models with heavy sterile neutrinos include cold dark matter candidates as
weakly interacting massive particles (WIMPs) \cite%
{Mizukoshi:2010ky,Dias:2010vt,Alvares:2012qv,Cogollo:2014jia}. A concise
review of WIMPs in 3-3-1 Electroweak Gauge Models is provided in Ref. \cite%
{daSilva:2014qba}.

In most versions of 3-3-1 models, one heavy triplet field with a Vacuum
Expectation Value (VEV) at a high energy scale breaks the symmetry $%
SU(3)_{L}\times U(1)_{X}$ into the SM electroweak group $SU(2)_{L}\times
U(1)_{Y}$, thus generating masses for the non SM fermions and non SM gauge
bosons, while other two lighter triplets with VEVs at the electroweak scale,
trigger the Electroweak Symmetry Breaking \cite{Hernandez:2013mcf} and
provide the masses for the SM particles. To provide an explanation for the
observed pattern of SM fermion masses and mixings, various 3-3-1 models with
flavor symmetries \cite%
{Kitabayashi:2002jd,Sen:2007vx,Dong:2010zu,Dong:2011vb,Vien:2013zra,Hernandez:2013hea,Vien:2014pta,Vien:2014gza,Vien:2014vka, Vien:2014ica,Hernandez:2014lpa,Hernandez:2014vta,VanVien:2015xha,Hernandez:2015tna,Hernandez:2015cra,Vien:2016tmh,Hernandez:2016eod}
and radiative seesaw mechanisms \cite%
{Kitabayashi:2000wh,Kitabayashi:2000nq,Kitabayashi:2001id,Kitabayashi:2002jd,Chang:2006aa,Dong:2006gx,Dong:2008sw, Martinez:2011iq,Hernandez:2013mcf,Huong:2012pg,Boucenna:2014ela,Okada:2015bxa}
have been proposed in the literature. However, some of them involve non
renormalizable interactions \cite%
{Hernandez:2014lpa,Hernandez:2014vta,Hernandez:2015tna,Hernandez:2015cra,Hernandez:2016eod}%
, others are renormalizable but do not address the observed pattern of
fermion masses and mixings due to the unexplained huge hierarchy among the
Yukawa couplings \cite%
{Dong:2010zu,Dong:2011vb,Vien:2013zra,Pires:2014xsa,Vien:2014pta,Vien:2014gza,Vien:2014vka,Vien:2014ica}
and others are only focused either in the quark mass hierarchy \cite%
{Dong:2006gx,Martinez:2011iq,Hernandez:2013hea}, or in the study of the
neutrino sector \cite%
{Kitabayashi:2000wh,Kitabayashi:2000nq,Kitabayashi:2001id,Kitabayashi:2002jd,Chang:2006aa,Dong:2008sw,Catano:2012kw,Boucenna:2014ela,Okada:2015bxa,Reig:2016ewy}%
, or only include the description of SM fermion mass hierarchy, without
addressing the mixings in the fermion sector \cite{Huong:2012pg}. It is
interesting to find an alternative explanation for the observed
SM fermion mass and mixing pattern, in the framework of 3-3-1 models, by
considering that it arises by a sequential loop suppression,
so that the masses are generated according to: three level top quark mass,
one loop level bottom, charm, tau and muon masses and two loop level masses
for the light up, down and strange quarks as well as for the electron and
neutrinos. This way of generating the SM fermion mass hierarchy was proposed
for the first time
in Ref. \cite{CarcamoHernandez:2016pdu}. However, the proposed model includes non-renormalizable Yukawa terms with a quite low cutoff scale.
In this paper we propose the first renormalizable extension of the 3-3-1
model with the electric charge constructed from the $SU(3)_{L}$ generators as $%
Q=T_{3}+\beta T_{8}+XI$ with $\beta =-\frac{1}{\sqrt{3}}$. The model
explains the SM fermion mass and mixing pattern by a sequential loop
suppression mechanism.

The paper is organized as follows. In section \ref{331model} we present the theoretical
setup of the proposed model. In section \ref{quarksection}
we discuss the quark masses and mixings within the model, while
the discussion of  the lepton masses and mixings
is given in section \ref{leptonmassesasndmixings}.

\section{The model}

\label{331model} The $SU(3)_{C}\times SU\left( 3\right) _{L}\times U\left(
1\right) _{X}$ model (3-3-1 model) with $\beta =-\frac{1}{\sqrt{3}}$ and
right-handed Majorana neutrinos in the $SU(3)_{L}$ lepton triplet was
proposed for the first time in \cite{Singer:1980sw}. However, the observed
pattern of fermion masses and mixings was not addressed at that time due to
the unexplained huge hierarchy among the Yukawa couplings \cite%
{Dong:2008sw,Pires:2014xsa,Dong:2010zu}.
Here we propose the first renormalizable extension of the 3-3-1 model with the parameter $%
\beta =-\frac{1}{\sqrt{3}}$, which includes a loop suppression mechanism to
generate the observed pattern of the SM fermion masses and mixings.
In our model only the top quark and the charged exotic fermions acquire tree
level masses, whereas the remaining SM fermions get their masses via
radiative corrections: 1 loop bottom, charm, tau and muon masses; 2-loop
masses for the light up, down, strange quarks as well as for the electron.
Light active neutrinos acquire their masses from a combination
of linear and inverse seesaw mechanisms at two loop level, and the quark
mixings arise from a combination of one and two loop level effects.

In order to realize this scenario we extend
the $SU(3)_{C}\times SU(3)_{L}\times U(1)_{X}$ group with an extra $Z_{4}\times Z_{2}$ discrete group, where the $Z_{4}$ symmetry is
softly broken and the remaining $Z_{2}$ symmetry is 
broken both spontaneously and softly. 
We also introduce a global $U(1)_{L_g}$ of the generalized lepton number
$L_{g}$ \cite{Chang:2006aa}, which is spontaneously broken down to a residual discrete $Z^{(L_g)}_{2}$ lepton number symmetry by a VEV of a gauge-singlet scalar
$\xi^{0}$ to be introduced below. The corresponding massless Goldstone boson, Majoron, is phenomenologically harmless being a gauge-singlet.
The full symmetry $\mathcal{G}$ of the model
experiences a two-step spontaneous breaking,
as follows:
\begin{eqnarray}
\mathcal{G}&=&
SU(3)_{C}\times SU\left( 3\right) _{L}\times U\left( 1\right)
_{X}\times Z_{4}\times Z_{2}\times U(1)_{L_g}\notag\\
&{\xrightarrow{v_{\chi },v_{\xi}}}&SU(3)_{C}\times
SU\left( 2\right) _{L}\times U\left( 1\right) _{Y}\times Z_4\times Z^{(L_g)}_{2}
\notag
\\
&&{\xrightarrow{v_{\eta }}}SU(3)_{C}\times U\left( 1\right) _{em}\times Z_{4}\times Z^{(L_g)}_{2}
,
\label{Group}
\end{eqnarray}
where the different symmetry breaking scales satisfy the following hierarchy
\begin{equation}
v_{\eta }=v=246\mbox{GeV}\ll v_{\chi }\sim v_{\xi}\sim \mathcal{O}(10)\mbox{TeV},
\label{eq:VEV-hierarchy}
\end{equation}
which corresponds in our model to the VEVs of the scalar fields to be
introduced below.
In the 3-3-1 model under consideration, the electric charge
is defined as \cite{Singer:1980sw,Valle:1983dk,Dong:2010zu}:
\begin{equation}
Q=T_{3}+\beta T_{8}+XI=T_{3}-\frac{1}{\sqrt{3}}T_{8}+XI,  \label{Q}
\end{equation}%
where $T_{3}$ and $T_{8}$ are the $SU(3)_{L}$ diagonal generators, $I$ is
the $3\times 3$ identity matrix and $X$ is the $U(1)_{X}$ charge.

Different versions of the $3\text{-}3\text{-}1$ models are determined by the
choice of the $\beta $ parameter, which is related to the different possible
fermion assignments. The most studied versions of $3\text{-}3\text{-}1$
models have $\beta =\pm \frac{1}{\sqrt{3}}$~\cite%
{Georgi:1978bv,Singer:1980sw} and $\beta =\pm \sqrt{3}$~\cite%
{Pisano:1991ee,Montero:1992jk,Frampton:1992wt}, and if we want to avoid
exotic charges we are led to only two different models: $\beta =\pm \frac{1}{\sqrt{3}}$.
Those having $\beta =\pm
\frac{1}{\sqrt{3}}$ contain non SM fermions with non-exotic electric
charges, i.e., equal to the electric charge of some SM fermions \cite{Sanchez:2001ua,Ponce:2001jn,Ponce:2002sg}. Those with $%
\beta =\pm \sqrt{3}$ have non SM fermions with large exotic electric charges
and require a departure from the perturbative regime at a scale of several
$\text{TeV}$, in order to successfully account for the measured value of the
weak mixing angle at low energies, as shown in detail in Ref. \cite%
{Martinez:2006gb}. Other versions of $3\text{-}3\text{-}1$ models have $%
\beta =0,\pm \frac{2}{\sqrt{3}}$ and contain non SM particles with fractional
electric charges \cite{Hue:2015mna}. For instance, $3\text{-}3\text{-}1$
models with $\ \beta =0$ contains exotic quarks and exotic charged leptons
with electric charges $\frac{1}{6}$ and $-\frac{1}{2}$, respectively \cite%
{Hue:2015mna}. Since electric charge conservation implies that the lightest
exotic particles of the $3\text{-}3\text{-}1$ models with $\beta =0,\pm 2/%
\sqrt{3}$ should be stable, the phenomenological viability of such models
requires a detailed analysis of the abundance of such stable exotic charged
particles in cosmology.

For these reasons, $3\text{-}3\text{-}1$ models with $\beta =-\frac{1%
}{\sqrt{3}}$ have advantages over those with $\beta =0,\pm \frac{2}{\sqrt{3}}%
,\pm \sqrt{3}$. In addition, choosing $\beta =-\frac{1}{\sqrt{3}}$ implies
that the third component of the weak lepton triplet is a neutral field $\nu
_{R}^{C}$, which allows building the Dirac matrix with the usual field $\nu
_{L}$ of the weak doublet. If one introduces a sterile neutrino $N_{R}$ in
the model, the light neutrino masses can be generated via low scale seesaw
mechanisms, which could be inverse or linear. The 3-3-1 models with $\beta =-%
\frac{1}{\sqrt{3}}$ can also provide an alternative framework to generate
neutrino masses, where the neutrino spectrum includes the light active
sub-eV scale neutrinos, as well as sterile neutrinos, which could be dark
matter candidates, if they are light enough, or candidates for detection at
the LHC, if their masses are at the TeV scale.
Therefore, pair production of TeV scale sterile neutrinos via the Drell-Yan
mechanism at the LHC could be a signal supporting models with extended gauge
symmetries such as the 3-3-1 models. In addition, Drell-Yan heavy
vector pair production processes at the LHC may help to distinguish the
3-3-1 models from other models with extended gauge symmetry.
With respect to the quark spectrum, we assign each of the first two families
of quarks to an $SU(3)_{L}$ antitriplet $3^{\ast }$, whereas the third
family is assigned to a $SU(3)_{L}$ triplet $3$, as required by the $%
SU(3)_{L}$ anomaly cancellation condition. Therefore, considering that there
are $3$ quark colors, we have six $3^{\ast }$ irreducible representations.
In addition, there are six $SU(3)_{L}$ triplets $3$ of fermionic fields,
considering the three lepton families. Thus, the $SU(3)_{L}$ representations
are vector like and anomaly free. The quantum numbers for the fermion
families are assigned in such a way that the combination of the $U(1)_{X}$
representations with other gauge sectors is anomaly free. As a consequence,
one finds that the number of chiral fermion generations is an integer
multiple of the number of colors, which provides an explanation for the
existence of three generations of quarks and leptons in
terms of the 3 colors. The $U(1)_{X}$-charge assignments of the fermionic
fields are obtained from Eq. (\ref{Q}) and the requirement of reproducing
the electric charges of the SM quarks and leptons. Then the $U(1)_{X}$
charge of the first two families of quark antitriplets is $X_{Q_{nL}}=\frac{1%
}{6}+\frac{\beta }{2\sqrt{3}}$ ($n=1,2$), whereas for the third family of
quark triplet is $X_{Q_{3L}}=\frac{1}{6}-\frac{\beta }{2\sqrt{3}}$, and
the corresponding $U(1)_{X}$-charges of the right handed quarks are equal to
their electric charges, given by $X_{u_{jR},d_{jR},J_{nR}}=\frac{2}{3%
},-\frac{1}{3},\frac{1}{6}+\frac{\sqrt{3}}{2}\beta $, ($j=1,2,3$ and $n=1,2$%
). The third generation non SM right handed quark $T_{R}$ has a $U(1)_{X}$%
-charge given by $X_{T_{R}}=\frac{1}{6}-\frac{\sqrt{3}}{2}\beta $. The three
left-handed lepton families are grouped into $SU(3)_{L}$ triplets with $%
X_{L_{jL}}=-\frac{1}{2}-\frac{\beta }{2\sqrt{3}}$ ($j=1,2,3$), while the
right-handed leptons are assigned as $SU(3)_{L}$ singlets with $U(1)_{X}$%
-charges equal to their electric charges, given by $X_{e_{iR},,\tilde{e}%
_{iR}}=-1,-\frac{1}{2}-\frac{\sqrt{3}}{2}\beta $, where $\tilde{e}_{iR}$ are
the right handed exotic leptons. These exotic fermions reside in vector-like
representations of the SM gauge group and are singlets under the $SU(2)_{L}$%
. Since we are considering a $3\text{-}3\text{-}1$ model with $\beta =-\frac{%
1}{\sqrt{3}}$, the cancellation of chiral anomalies implies that quarks are
unified in the following $SU(3)_{C}\times SU(3)_{L}\times U(1)_{X}$ left-
and right-handed representations \cite%
{Valle:1983dk,Hoang:1995vq,Diaz:2004fs,CarcamoHernandez:2005ka}:
\begin{align}
Q_{nL}& =%
\begin{pmatrix}
D_{n} \\
-U_{n} \\
J_{n} \\
\end{pmatrix}%
_{L}\sim \left( 3,3^{\ast },0\right) ,\hspace{1cm}Q_{3L}=%
\begin{pmatrix}
U_{3} \\
D_{3} \\
T \\
\end{pmatrix}%
_{L}\sim \left( 3,3,\frac{1}{3}\right) ,\hspace{1cm}n=1,2,  \notag \\
D_{iR}& \sim \left( 3,1,-\frac{1}{3}\right) ,\hspace{1cm}U_{iR}\sim \left(
3,1,\frac{2}{3}\right),\hspace{1cm}i=1,2,3,\notag\\
J_{nR}&\sim \left( 3,1,-\frac{1}{3}%
\right) ,\hspace{1cm}T_{R}\sim \left( 3,1,\frac{2}{3}\right),%
\end{align}
where $U_{iL}$ and $D_{iL}$ ($i=1,2,3$) are the left handed up and down type
quarks fields in the flavor basis, respectively. The right handed SM\
quarks, i.e., $U_{iR}$ and $D_{iR}$ ($i=1,2,3$) and right handed exotic
quarks, i.e., $T_{R}$ and $J_{nR}$ ($n=1,2$)\ are assigned to be $SU(3)_{L}$
singlets with $U(1)_{X}$ quantum numbers equal to their electric charges.

Furthermore, the requirement of chiral anomaly cancellation constrains the
leptons to the following \mbox{$SU(3)_{C}\times SU(3)_{L}\times U(1)_{X}$}
left- and right-handed representations \cite%
{Valle:1983dk,Hoang:1995vq,Diaz:2004fs}:
\begin{equation}
L_{iL}=%
\begin{pmatrix}
\nu _{i} \\
e_{i} \\
\nu _{i}^{c} \\
\end{pmatrix}%
_{L}\sim \left( 1,3,-\frac{1}{3}\right) ,\hspace{1cm}e_{iR}\sim (1,1,-1),%
\hspace{1cm}i=1,2,3,  \label{L}
\end{equation}%
where $\nu _{iL}, \nu^{c} \equiv \nu_{R}^{c}$ and $e_{iL}$ ($e_{L},\mu
_{L},\tau _{L}$) are the neutral and charged lepton families, respectively.
Let us note that we assign the right-handed leptons to $SU(3)_{L}$ singlets,
which implies that their $U(1)_{X}$ quantum numbers correspond to their
electric charges.

To implement the radiative seesaw mechanisms that generate the observed
hierarchy of the SM charged fermion masses and mixing angles by a sequential
loop suppression and the light active neutrino masses from a combination of
linear and inverse seesaw mechanisms at two loop level, we extend both the
fermion and the scalar sectors of the 3-3-1 models with $\beta =-\frac{1}{%
\sqrt{3}}$ previously considered in the literature. We introduce $SU(3)_{L}$ singlet exotic up type quarks $\widetilde{T}_{L,R}$, down type quarks $B_{L,R}$ and charged leptons $E_{L,R}$ as well as four gauge group Eq.~(\ref{Group}) singlet leptons $N_{R},\Psi _{R}$. Their complete $%
SU(3)_{C}\times SU(3)_{L}\times U(1)_{X}$ assignments are:%
\begin{eqnarray}
\widetilde{T}_{1L} &\sim &(3,1,2/3),\hspace{1cm}\widetilde{T}_{1R}\sim
(3,1,2/3),\hspace{1cm}\widetilde{T}_{2L}\sim (3,1,2/3),\hspace{1cm}%
\widetilde{T}_{2R}\sim (3,1,2/3),  \notag \\
B_{L} &\sim &(3,1,-1/3),\hspace{1cm}B_{R}\sim (3,1,-1/3), \\
E_{1L} &\sim &\left( 1,1,-1\right) ,\hspace{1cm}E_{2L}\sim \left(
1,1,-1\right) ,\hspace{1cm}E_{3L}\sim \left( 1,1,-1\right) ,  \notag \\
E_{1R} &\sim &\left( 1,1,-1\right) ,\hspace{1cm}E_{2R}\sim \left(
1,1,-1\right) ,\hspace{1cm}E_{3R}\sim \left( 1,1,-1\right) ,  \notag \\
N_{1R} &\sim &\left( 1,1,0\right) ,\hspace{1cm}N_{2R}\sim \left(
1,1,0\right) ,\hspace{1cm}N_{3R}\sim \left( 1,1,0\right) ,\hspace{1cm}\Psi
_{R}\sim \left( 1,1,0\right) .
\end{eqnarray}
The $U(1)_{L_g}\times Z_{4}\times Z_{2}$ assignments for all the fermions of the model are
shown in \mbox{Tables~\ref{ta:quarks}, \ref{ta:leptons}.}

\begin{table}[tbp]
\resizebox{16cm}{!}{
\renewcommand{\arraystretch}{1.2}
\begin{tabular}{|c|c|c|c|c|c|c|c|c|c|c|c|c|c|c|c|c|c|c|}
\hline
 & $Q_{1L}$ & $Q_{2L}$ & $Q_{3L}$ & $U_{1R}$ & $U_{2R}$ & $U_{3R}$ & $T_R$ & $D_{1R}$ & $D_{2R}$ & $D_{3R}$ & $J_{1R}$ & $J_{2R}$ & $\widetilde{T}_{1L}$ & $\widetilde{T}_{1R}$ & $\widetilde{T}_{2L}$ & $\widetilde{T}_{2R}$ & $B_L$ & $B_R$ \\ \hline
$L_g$ & $\frac{2}{3}$ & $\frac{2}{3}$ & $-\frac{2}{3}$ & $0$ & $0$ & $0$ & $-2$ & $0$ & $0$ & $0$ & $2$ & $2$ & $0$ & $0$ & $0$ & $0$ & $0$ & $0$ \\ \hline$$
 $Z_4$ & $-1$ & $-1$ & $1$ & $1$ & $-i$ & $1$ & $1$ & $1$ & $1$ & $1$ & $-1$ & $-1$ & $i$ & $1$ & $i$ & $1$ & $-1$ & $-1$ \\ \hline
  $Z_2$ & $1$ & $1$ & $1$ & $1$ & $1$ & $-1$ & $-1$ & $1$ & $1$ & $1$ & $-1$ & $-1$ & $1$ & $1$ & $1$ & $1$ & $1$ & $1$ \\ \hline
\end{tabular}}
\caption{Quark assignments under
$Z_{4}\times Z_{2}$ and the values of generalized Lepton Number $L_{g}$.
}
\label{ta:quarks}
\end{table}
\begin{table}[tbp]
\resizebox{16cm}{!}{
\renewcommand{\arraystretch}{1.2}
\begin{tabular}{|c|c|c|c|c|c|c|c|c|c|c|c|c|c|c|c|c|}
\hline
 & $L_{1L}$ & $L_{2L}$ & $L_{3L}$ & $e_{1R}$ & $e_{2R}$ & $e_{3R}$ & $E_{1L}$ & $E_{2L}$ & $E_{3L}$ & $E_{1R}$ & $E_{2R}$ & $E_{3R}$ & $N_{1R}$ & $N_{2R}$ & $N_{3R}$ & $\Psi_{R}$ \\ \hline
 $L_g$ & $\frac{1}{3}$ & $\frac{1}{3}$ & $\frac{1}{3}$ & $1$ & $1$ & $1$ & $1$ & $1$ & $1$ & $1$ & $1$ & $1$ & $-1$ & $-1$ & $-1$ & $1$ \\ \hline$$
 $Z_4$ & $i$ & $i$ & $i$ & $-i$ & $-i$ & $-i$ & $1$ & $i$ & $i$ & $-i$ & $-i$ & $-i$ & $i$ & $i$ & $i$ & $1$ \\ \hline
  $Z_2$ & $-1$ & $1$ & $1$ & $-1$ & $1$ & $1$ & $-1$ & $1$ & $1$ & $-1$ & $1$ & $1$ & $-1$ & $-1$ & $-1$ & $-1$ \\ \hline
\end{tabular}}
\caption{
Lepton assignments under $Z_{4}\times Z_{2}$ and the values of generalized Lepton Number $L_{g}$.
}
\label{ta:leptons}
\end{table}

\begin{table}[tbp]
\begin{tabular}{|c|c|c|c|c|c|c|c|c|c|c|}
\hline
 & $\chi$ & $\eta$ & $\rho$ & $\varphi_{1}^{0}$ & $\varphi_{2}^{0}$ & $\phi_{1}^{+}$ & $\phi_{2}^{+}$ & $\phi_{3}^{+}$ & $\phi_{4}^{+}$ & $\xi^{0}$\\ \hline
 $L_g$ & $\frac{4}{3}$ & $-\frac{2}{3}$ & $-\frac{2}{3}$ & $0$ & $0$ & $0$ & $-2$ & $-2$ & $-2$ & $-2$ \\ \hline
$Z_4$ & $1$ & $1$ & $-1$ & $-1$ & $i$ & $i$ & $-1$ & $-1$ & $1$ & $1$ \\ \hline
$Z_2$ & $-1$ & $-1$ & $1$ & $1$ & $1$ & $1$ & $1$ & $-1$ & $-1$ & $1$\\ \hline
\end{tabular}\vspace{-0.1cm}
\caption{Scalar assignments under
$Z_{4}\times Z_{2}$ and the values of generalized Lepton Number $L_{g}$.
}
\label{ta:scalars}
\end{table}

Compared to the simplified versions of 3-3-1 models with the scalar sector
composed only of three $SU(3)_{L}$ scalar triplets -- $\chi $, $\eta $ and $%
\rho $ -- we introduce seven $SU(3)_{L}$ singlets $\varphi _{1}^{0}$,$\
\varphi _{2}^{0}$, $\xi^{0}$,$\ \phi _{1}^{+}$,$\ \phi _{2}^{+}$, $\phi _{3}^{+}$ and $%
\phi _{4}^{+}$. All these scalars are assigned in our model to the following
representations of $SU(3)_{C}\times SU(3)_{L}\times U(1)_{X}$:%
\begin{align}
\chi & =%
\begin{pmatrix}
\chi _{1}^{0} \\
\chi _{2}^{-} \\
\frac{1}{\sqrt{2}}(v_{\chi }+\xi _{\chi }\pm i\zeta _{\chi })%
\end{pmatrix}%
\sim \left( 1,3,-\frac{1}{3}\right) ,\hspace{1cm}\rho =%
\begin{pmatrix}
\rho _{1}^{+} \\
\frac{1}{\sqrt{2}}(\xi _{\rho }\pm i\zeta _{\rho }) \\
\rho _{3}^{+}%
\end{pmatrix}%
\sim \left( 1,3,\frac{2}{3}\right) ,  \notag \\
\eta & =%
\begin{pmatrix}
\frac{1}{\sqrt{2}}(v_{\eta }+\xi _{\eta }\pm i\zeta _{\eta }) \\
\eta _{2}^{-} \\
\eta _{3}^{0}%
\end{pmatrix}%
\sim \left( 1,3,-\frac{1}{3}\right) ,\hspace{1cm}\varphi _{1}^{0}\sim
(1,1,0),\hspace{1cm}\varphi _{2}^{0}\sim (1,1,0),  \notag \\
\phi _{1}^{+}& \sim (1,1,1),\hspace{0.5cm}\phi _{2}^{+}\sim (1,1,1),\hspace{0.5cm}\phi _{3}^{+}\sim (1,1,1),\hspace{0.5cm}\phi _{4}^{+}\sim (1,1,1),\hspace{0.5cm}\xi^{0}\sim (1,1,0),
\label{scalarsector1}
\end{align}
Their $U(1)_{L_g}\times Z_{4}\times Z_{2}$ assignments are shown in Table \ref{ta:scalars}.

The spontaneous symmetry breaking (\ref{Group}) in our model is triggered by
the VEVs (\ref{eq:VEV-hierarchy}) of the scalar fields $\chi $, $\eta $ and $\xi^{0}$,
neutral under the $Z_{4}$ discrete symmetry. As seen from (\ref{Group}), the
first stage of the breaking is done by a TeV scale VEV $v_{\chi }$ of an
$SU(3)_{L}$ triplet $\chi $ handing masses to the non-SM fermions and gauge
bosons
as well as by the TeV scale VEV $v_{\xi}$ of the gauge-singlet scalar $\xi^0$, which spontaneously
 breaks the generalized lepton number symmetry $U(1)_{L_g}$.
The corresponding Majoron is a gauge-singlet and, therefore, unobservable.
Note that Lepton Number (LN) is broken together with Generalized Lepton Number (GLN) by the VEV of
$\xi^0$, which has both LN and GLN equal to $-2$.
Since the gauge singlet scalar $\xi^{0}$ breaks $U(1)_{L_g}$
in a way that respects the condition
$\left|\Delta L_g\right|=\left|\Delta L\right|=2$, there survives a residual discrete $Z^{(L_g)}_{2}$ lepton number symmetry
under which the leptons are charged and the other particles are neutral. This means that in any reaction leptons can appear only in pair, thus, forbidding proton decay.
 The TeV scale VEVs $v_{\chi }$ of the $SU(3)_{L}$ triplet $\chi $ also breaks $Z_{2}$ symmetry.
 Another $SU(3)_{L}$ triplet $\eta $ with a Fermi
scale VEV $v_{\eta }$ is responsible for the electroweak symmetry breaking
and the masses of the SM fermions and $W,Z$-bosons.

Let us explain the VEV pattern of the $SU(3)_{L}$ scalar triplets $\chi $
and $\eta $.
Since the 
$\chi $ triggers the $SU(3)_{L}\times U(1)_{X}\rightarrow SU(2)_{L}\times
U(1)_{Y}$ breaking, the following conditions have to be fulfilled: \
\begin{equation}
T_{1}\left\langle \chi \right\rangle =T_{2}\left\langle \chi \right\rangle
=T_{3}\left\langle \chi \right\rangle =\left( \beta T_{8}+XI\right)
\left\langle \chi \right\rangle =0.  \label{chiVEVconditions}
\end{equation}%
whereas the remaining generators do not leave the vacuum $\left\langle \chi
\right\rangle $ invariant. From the first three conditions for $\left\langle
\chi \right\rangle $ given in Eq. (\ref{chiVEVconditions}), it follows that:
\begin{equation}
\left\langle \chi \right\rangle =%
\begin{pmatrix}
0 \\
0 \\
\frac{v_{\chi }}{\sqrt{2}}%
\end{pmatrix}
\label{tbchchi}
\end{equation}%
The last condition in Eq. (\ref{chiVEVconditions}) for $\left\langle \chi
\right\rangle $, i.e, $\left( \beta T_{8}+XI\right) \left\langle \chi
\right\rangle =0$, yields the following relation between the $U\left(
1\right) _{X}$ charge of the $SU(3)_{L}$ scalar triplet $\chi $ and the $%
\beta $ parameter:
\begin{equation}
X_{\chi }=\frac{\beta }{\sqrt{3}},
\end{equation}%
which for $\beta =-\frac{1}{\sqrt{3}}$, results in $X_{\chi }=-\frac{1}{3}$,
as indicated by Eq. (\ref{scalarsector1}).

The electroweak symmetry breaking $SU(2)_{L}\times U(1)_{Y}\rightarrow
U(1)_{EM}$ in our model is realized by the VEV $v_{\eta }$ of the $SU(3)_{L}$
scalar triplet $\eta $. Requiring that all the $SU\left( 3\right) _{L}$
generators are broken, with the exception of the electric charge generator $Q$, we arrive at the following VEV pattern
\begin{equation}
\left\langle \eta \right\rangle =%
\begin{pmatrix}
\frac{v_{\eta }}{\sqrt{2}} \\
0 \\
0%
\end{pmatrix}%
.  \label{eq:eta-VEV}
\end{equation}%
From the requirement of the $U(1)_{EM}$ invariance we have
\begin{equation}
Q\left\langle \eta \right\rangle =\left( T_{3}+\beta T_{8}+XI\right)
\left\langle \eta \right\rangle =0,  \label{eq:Q-conservation-condition}
\end{equation}%
thus producing the following relation for the $U(1)_{X}$ charge $X_{\eta }$
of the $\eta $ field:
\begin{equation}
X_{\eta }=-\frac{1}{2}-\frac{\beta }{2\sqrt{3}},
\end{equation}%
which for $\beta =-\frac{1}{\sqrt{3}}$ results in $X_{\eta }=-\frac{1}{3}$
as indicated by Eq. (\ref{scalarsector1}).

Note that, the difference between the $\eta$ and $\chi$ Higgs triplets can be explained using the generalized lepton number $L_g$, discussed in Appendix \ref{gln}.
Its values for the fields of the model are specified in Tables~\ref{ta:quarks}-\ref{ta:scalars}.

The choice of the VEV structure in (\ref{tbchchi}) and (\ref{eq:eta-VEV}) shows that only the neutral Higgs field without lepton number is allowed to have the VEV. In addition, the patterns of the $SU(3)_L$ scalar triplets $\chi$ and $\eta$ shown in Eq. (\ref{tbchchi}) and (\ref{eq:eta-VEV}) are consistent with a global minimum of the scalar potential of our model for all the region of parameter space.
We adopt $X_{\rho }=2/3$ for another $SU(3)_{L}$ scalar triplet $\rho $ in
Eq. (\ref{scalarsector1}) from the simplified versions of the 3-3-1 model
\cite{Foot:1994ym,Hoang:1995vq,Singer:1980sw},
where both $\eta $ and $\rho $ scalars participate in the electroweak
symmetry breaking. The extra $SU(3)_{L}$ scalar triplet $\rho$ is introduced in simplified versions of the 3-3-1 models to give masses to charged leptons, as well as to the bottom, up and charm quarks. In our model the $SU(3)_L$ scalar triplet $\rho$ is crucial to give one loop level masses for the bottom and charm quarks, to the tau and muon leptons as well as two loop level masses for the up, down and strange quarks as well to the electron, as shown in Figs.~\ref{LoopdiagramsUD}, \ref{Loopdiagrams3}.
The $SU(3)_{L}$ scalar triplet $\rho$ also contributes to some
entries of the neutrino mass matrix as indicated in Fig.~\ref{Loopdiagrams2}.
On the other hand, the conditions similar to (\ref{eq:eta-VEV}), (\ref%
{eq:Q-conservation-condition}) are applied to $\langle \rho \rangle $ as
well and lead to $X_{\rho }=2/3$. In our model we have $\langle \rho \rangle
=0$ due to the $Z_{4}$ conservation (\ref{Group}), and the above symmetry
breaking conditions do not restrict $X_{\rho }$.
We choose $X_{\rho }=2/3$ in order to maintain resemblance with the previous
versions of the 3-3-1 model. Another motivation for the choice $X_{\rho
}=2/3 $ is the $U(1)_{X}$ invariance of the $SU(3)_{L}$ invariant trilinear
scalar interaction $\chi \eta \rho $. Let us note that our choice $\beta =-%
\frac{1}{\sqrt{3}}$ yields $X_{\eta }=X_{\chi }=-\frac{1}{3}$, which in turn
leads to $X_{\rho }=2/3$.

With the above particle content, the relevant quark and lepton Yukawa terms
invariant under the
symmetry group (\ref{Group}) of our model
take the form:
\begin{eqnarray}
-L_{gY}^{\left( q\right) } &=&
h_{\chi }^{\left( T\right) }\overline{Q}_{3L}\chi T_{R}+
h_{\eta }^{\left( U\right) }\overline{Q}_{3L}\eta U_{3R}
\notag \\
&&+\sum_{n=1}^{2}\sum_{m=1}^{2}h_{\rho nm}^{\left( \widetilde{T}\right) }%
\overline{Q}_{nL}\rho ^{\ast }\widetilde{T}_{mR}+\sum_{n=1}^{2}h_{\varphi
_{1}^{0}n2}^{\left( U\right) }\overline{\widetilde{T}}_{nL}\varphi
_{1}^{0}U_{2R}+\sum_{n=1}^{2}h_{\varphi _{2}^{0}n1}^{\left( U\right) }%
\overline{\widetilde{T}}_{nL}\varphi _{2}^{0}U_{1R}  \notag \\
&&+\sum_{n=1}^{2}\sum_{m=1}^{2}h_{\chi nm}^{\left( J\right) }\overline{Q}%
_{nL}\chi ^{\ast }J_{mR}+h_{\rho }^{\left( B\right) }\overline{Q}_{3L}\rho
B_{R}+\sum_{j=1}^{3}h_{\varphi^{0}_{1}j}^{\left( D\right) }{\overline{B}_{L}}%
\varphi _{1}^{0}D_{jR}\\
&&+\sum_{n=1}^{2}\sum_{j=1}^{3}h_{\phi
_{1}^{+}nj}^{\left( D\right) }\overline{\widetilde{T}}_{nL}\phi
_{1}^{+}D_{jR}+\sum_{n=1}^{2}\sum_{m=1}^{2}h_{\varphi _{2}^{0}nm}^{\left(
\widetilde{T}\right) }\overline{\widetilde{T}}_{nL}\varphi _{2}^{0}%
\widetilde{T}_{mR} +  {m_{B} \bar{B}_{L} B_{R}} +h.c, \notag \label{Lyquarks}\\[3mm]
-L_{gY}^{\left( l\right) } &=&h_{\rho }^{\left( E\right) }\overline{%
L}_{1L}\rho E_{1R}+h_{\varphi^{0}_{2}}^{\left( E\right) }\overline{E}%
_{1L}\varphi _{2}^{0}E_{1R}+h_{\varphi^{0}_{2}}^{\left( e\right) }\overline{E%
}_{1L}\varphi _{2}^{0}e_{1R}+\sum_{n=2}^{3}\sum_{m=2}^{3}h_{\rho nm}^{\left(
E\right) }\overline{L}_{nL}\rho E_{mR}  \notag \\
&& + h_{\rho }^{\left(e\right) }\overline{L}_{1L}\rho e_{1R}+ \sum_{n=2}^{3}\sum_{m=2}^{3}h_{\rho nm}^{\left(e\right) }\overline{L}_{nL}\rho e_{mR}  + \sum_{n=2}^{3}\sum_{m=2}^{3}h_{\varphi^{0}_{1}nm}^{\left(E\right) }\overline{E}_{nL}\varphi _{1}^{0}E_{mR}\notag \\
&&+\sum_{n=2}^{3}\sum_{m=2}^{3}h_{\varphi^{0}_{1}nm}^{\left( e\right) }%
\overline{E}_{nL}\varphi _{1}^{0}e_{mR}
+\sum_{n=2}^{3}\sum_{j=1}^{3}h_{\chi nj}^{\left( L\right) }\overline{L}%
_{nL}\chi N_{jR} \notag\\
&&+\sum_{j=1}^{3}\sum_{n=2}^{3}h_{\phi _{4}^{-}nj}^{\left( e\right) }%
\overline{E}_{nL}\phi
_{4}^{-}N_{jR}+\sum_{j=1}^{3}h_{\varphi^{0}_{2}}^{\left( N\right)
}\overline{\Psi _{R}^{c}}\left( {\varphi _{2}^{0}}\right) ^{\ast }N_{jR}
+ y_{\Psi}\overline{\Psi _{R}^{c}}\Psi _{R}\xi^0\notag\\
&&+h_{\rho 11}^{\left( L\right) }\varepsilon _{abc}\overline{L}_{1L}^{a}\left(L_{1L}^{C}\right)^{b}\left( \rho ^{\ast }\right) ^{c}+\sum_{n=2}^{3}\sum_{m=2}^{3}h_{\rho nm}^{\left( L\right) }\varepsilon _{abc}\overline{L}_{nL}^{a}\left(L_{mL}^{C}\right)^{b}\left( \rho ^{\ast }\right) ^{c}+h.c.
\label{Lyleptons}
\end{eqnarray}
where the dimensionless parameters in Eqs. (\ref{Lyquarks}) and (\ref{Lyleptons}) are $%
\mathcal{O}(1)$ dimensionless couplings. From the quark Yukawa terms it
follows that the top quark mass mainly arises from the interaction with the $%
SU(3)_{L}$ scalar triplet $\eta $, which breaks the $SU\left( 2\right)
_{L}\times U\left( 1\right) _{Y}$ gauge group. Consequently, the dominant
contribution to the SM-like $126$ GeV Higgs boson arises mainly from the CP
even neutral component $\xi _{\eta }$ of the $SU(3)_{L}$ scalar triplet $\eta
$. The terms of the scalar potential relevant for the implementation of the
radiative seesaw mechanisms that generate the observed hierarchy of the SM
charged fermion masses and mixing angles by a sequential loop suppression
are:
\begin{equation}
V\supset \lambda _{1}\eta \chi \rho \varphi _{1}^{0}+\lambda _{2}\eta \chi
\rho \left( \varphi _{1}^{0}\right) ^{\ast }+\lambda_{3}\phi _{3}^{-}\rho \eta^{\dagger }\xi^0+
\lambda_{4}\phi _{1}^{-}\phi _{2}^{+}\left( \varphi _{2}^{0}\right) ^{\ast
}\left(\xi^{0}\right)^{\ast}+w_{1}\left( \varphi_{2}^{0}\right)^{2}\varphi _{1}^{0}+w_{2}\phi _{3}^{-}\rho \chi ^{\dagger }+h.c.\,.  \label{relscalarterms}
\end{equation}
After the spontaneous breaking of the electroweak symmetry, the above-given
Yukawa interactions generate the observed hierarchy of SM fermion masses and
mixing angles by a sequential loop suppression, provided that one introduces
the $Z_{4}\times Z_{2}$ soft breaking mass terms for the electroweak singlet
fermions:
\begin{eqnarray}
L_{g soft}^{F}&=&\sum_{n=1}^{2}\sum_{m=1}^{2}\left( m_{\widetilde{T}%
}\right) _{nm}\overline{\widetilde{T}}_{nL}\widetilde{T}_{mR}+m_{E_{1}}%
\overline{E}_{1L}E_{1R}+\sum_{n=2}^{3}\sum_{m=2}^{3}\left( m_{E}\right) _{nm}%
\overline{E}_{nL}E_{mR}\notag\\
&&+\sum_{n=2}^{3}\left( m_{E}\right) _{n1}\overline{E}%
_{nL}E_{1R}+h.c., \label{eq:Fermion-Soft-terms}
\end{eqnarray}
as well as soft $Z_{4}\times Z_{2}$ breaking in the electroweak singlet
scalar sector:
\begin{equation}
L_{g soft}^{scalars}=\mu _{1}^{2}\left( \varphi _{2}^{0}\right)
^{2}+\mu _{2}^{2}\phi _{2}^{-}\phi _{3}^{+}+\mu _{3}^{2}\phi _{4}^{-}\phi
_{3}^{+}+h.c.\,.  \label{eq:Scalar-Soft-terms}
\end{equation}

Let us note that in the simplified version of the 3-3-1 model with $\beta =-%
\frac{1}{\sqrt{3}}$, whose scalar sector contains three $SU(3)_{L}$ scalar
triplets, the flavor constraints can be fulfilled by considering the scale
of breaking of the $SU(3)_{L}\times U(1)_{X}$ gauge symmetry much larger
than the electroweak symmetry breaking scale $v=246$ GeV, which corresponds
to the alignment limit of the mass matrix for the CP-even Higgs bosons \cite{Okada:2016whh}. Our model has a more extended scalar sector since it is
composed of three $SU(3)_{L}$ scalar triplets (from which one is inert $%
SU(3)_{L}$ triplet) and six $SU(3)_{L}$ scalar singlets. Consequently,
following Ref. \cite{Okada:2016whh}, we expect that the FCNC effects as well
as the constraints arising from $K^{0}-\bar{K^{0}}$, $B^{0}-\bar{B^{0}}$ and
$D^{0}-\bar{D^{0}}$ mixings will be fulfilled in our model, by considering
the scale of breaking of the $SU(3)_{L}\times U(1)_{X}$ gauge symmetry much
larger than the scale of breaking of the electroweak symmetry.
The scalar sector of our model is not predictive as its corresponding scalar
potential has many free uncorrelated parameters that can be adjusted to get
the required pattern of scalar masses. Therefore, the loop effects of the
heavy scalars contributing to certain observables can be suppressed by the
appropriate choice of the free parameters in the scalar potential.
Fortunately, all these adjustments do not affect the charged fermion and
neutrino sector, which is completely controlled by the fermion-Higgs Yukawa
couplings.

Despite the fact that the scalar and fermion sectors of our model are
considerably larger than the corresponding to the simplified version of the
3-3-1 model with $\beta =-\frac{1}{\sqrt{3}}$, and the fields assignments
under the discrete group $Z_{4}\times Z_{2}$ look rather sophisticated, each
introduced element plays its own role in the implementation of the radiative
seesaw mechanisms that allow us to explain the SM fermion mass hierarchy by
a sequential loop suppression. In what follows we provide a justification
and summary of our above presented model setup:

\begin{enumerate}
\item The spontaneously and softly broken $Z_{2}$ symmetry is crucial to
generate a two loop level electron mass as it distinguishes the first
generation left leptonic triplet, i.e., $L_{1L}$, neutral under $Z_{2}$ from
the second and third generation ones i.e., $L_{2L}$\ and $L_{3L}$ which are $%
Z_{2}$ even. This symmetry also separates the SM right handed charged
leptonic field, i.e, $e_{1R}$, which is $Z_{2}$ odd from the remaining SM
right handed charged leptonic fields, i.e, $e_{2R}$ and $e_{3R}$, neutral
under the $Z_{2}$ symmetry. This results in one loop level tau and muon
lepton masses and a two loop level mass for the electron.

\item The softly broken $Z_{4}$ symmetry separates the third generation left
handed quark fields from the first and second generation ones, giving rise
to a tree level top and exotic quark masses and to radiatively generated
masses for the remaining quarks. Besides that, the $Z_{4}$ symmetry
differentiates the second generation right handed SM up quark fields, i.e., $%
U_{2R}$, charged under this symmetry, from the first generation SM one,
i.e., $U_{1R}$, which is $Z_{4}$ neutral, thus giving rise to a one loop
level charm quark mass and two loop level up quark mass.

\item The scalar sector of our model is composed of three $SU(3)_{L}$ scalar
triplets, i.e., $\chi $, $\eta $ and $\rho $, seven $SU(3)_{L}$ scalar
singlets, from which three are electrically neutral, i.e., $\varphi _{1}^{0}$, $\ \varphi _{2}^{0}$ and $\xi^{0}$ and four electrically charged, i.e.,$\ \phi _{1}^{+}$
,$\ \phi _{2}^{+}$,$\ \phi _{3}^{+}$ and $\phi _{4}^{+}$. The inclusion of
the spontaneously and softly broken $Z_{2}$ symmetry requires the
introduction of a $SU(3)_{L}$ scalar singlet $\phi _{3}^{+}$, which is odd
under this symmetry. The presence of the $SU(3)_{L}$ scalar singlet $\phi
_{3}^{+}$, is needed in order to build the $Z_{2}$ invariant trilinear
scalar interactions required to generate two loop level down and strange
quark masses, as shown in Fig.~\ref{LoopdiagramsUD}. Besides that, in order
to implement a two loop level radiative seesaw mechanism for the generation
of the up, down and strange quark masses as well as the electron mass, the $%
Z_{4}$ charged $SU(3)_{L}$ scalar singlets $\varphi _{1}^{0}$,$\ \varphi
_{2}^{0}$, $\phi _{1}^{+}$,$\ \phi _{2}^{+}$ (which do not acquire a vacuum
expectation value) are also required in the scalar sector. The $Z_{4}$
charged $SU(3)_{L}$ scalar singlet $\varphi _{1}^{0}$ is also needed for the
implementation of the one loop level radiative seesaw mechanism that
generates the charm, the bottom quark masses as well as the tau and muon
lepton masses, as shown in Fig.~\ref{Loopdiagrams3}. The $Z_{4}$ charged $%
SU(3)_{L}$ scalar singlets $\varphi _{2}^{0}$ and $\phi _{3}^{+}$ as well as
the $SU(3)_{L}$ scalar singlet $\phi _{4}^{+}$, neutral under $Z_{4}$ are
also crucial for the implementation of two loop level linear and inverse
seesaw mechanisms that give rise to the light active neutrino masses. The $SU(3)_{L}$ scalar
singlet $\xi^{0}$ is introduced to spontaneously break the $U(1)_{L_g}$ generalized lepton number symmetry and thus giving rise to a tree-level mass for the right handed Majorana neutrino $\Psi_R$.  Lets us note that $\xi^{0}$ is the only electrically neutral $SU(3)_{L}$ scalar
singlet that has a non-vanishing generalized Lepton Number $L_g$.
It is crucial for generating two loop-level masses for the down and strange quarks. This is
due to the fact that the electrically charged $SU(3)_L$ scalar singlets $\phi^+_2$ and $\phi^+_3$ appearing in the two loop level diagrams that give rise to the down and strange quark masses carry non-vanishing generalized Lepton Numbers thus implying that the quartic scalar interaction $\lambda_{3}\phi _{3}^{-}\rho \eta^{\dagger }\xi^0$ is crucial to generate the masses for the down and strange quarks, as shown in Fig.~\ref{LoopdiagramsUD}. Note that we assign non-vanishing generalized Lepton Numbers for $\phi^+_2$ and $\phi^+_3$ because $\phi^+_3$ mix with $\phi^+_4$ as well as with $\phi^+_2$ via the soft breaking mass terms of Eq. (\ref{eq:Scalar-Soft-terms}) and $\phi^+_4$ carry a non-vanishing generalized Lepton Number as required from the invariance of the lepton Yukawa interaction $\overline{E}_{nL}\phi_{4}^{-}N_{jR}$ under the $U(1)_{L_g}$ symmetry.
\item The fermion sector of the 3-3-1 model, with right-handed %
neutrinos $\nu _{R}^{c}$ in the $SU(3)_{L}$ lepton triplet, is extended by
introducing two $SU(3)_{L}$ singlet exotic up type quarks, i.e. $\widetilde{T%
}_{1}$ and $\widetilde{T}_{2}$, a $SU(3)_{L}$ singlet exotic down type
quark, i.e., $B$, three $SU(3)_{L}$ singlet exotic charged leptons, i.e., $%
E_{j}$ ($j=1,2,3$) and four right handed Majorana neutrinos $N_{jR}$ ($%
j=1,2,3$), $\Psi _{R}$. The $SU(3)_{L}$ singlet exotic down type quarks,
i.e. $B$, is crucial for the implementation of the one loop level radiative
seesaw mechanism that generate the bottom quark mass. The $SU(3)_{L}$
singlet exotic up type quarks, i.e., $\widetilde{T}_{1}$ and $\widetilde{T}%
_{2}$, are needed to generate a one loop level charm quark mass as well as
two loop level down and strange quark masses. The three $SU(3)_{L}$ singlet
exotic charged leptons, i.e., $E_{j}$ ($j=1,2,3$), are required in order to
provide the radiative seesaw mechanisms that generate one loop level tau
and muon masses and two loop level electron mass. The four right handed
Majorana neutrinos, i.e., $N_{jR}$ ($j=1,2,3$), $\Psi _{R}$, are crucial for
the implementation of the two loop level linear and inverse seesaw
mechanisms that give rise to the light active neutrino masses. It is worth
mentioning that out of these four right handed Majorana neutrinos, only $%
\Psi _{R}$ acquires a tree level mass, whereas the three remaining right
handed Majorana neutrinos, i.e., $N_{jR}$ ($j=1,2,3$), get their masses via a
one loop level radiative seesaw mechanism mediated by $\Psi _{R}$ and $%
\varphi _{2}^{0}$, as shown in Fig.~\ref{Loopdiagrams2}.
\end{enumerate}

In what follows we briefly comment on some phenomenological
aspects of our model concerning LHC signals of non-SM fermions. From the
quark Yukawa interactions it follows that the heavy exotic $SU(3)_{L}$
singlet down (up) type quark(s), i.e., $B$ ($\widetilde{T}_{n}$ ($n=1,2$))
will decay predominantly into a SM down (up) type quark and the $Re%
\varphi _{1}^{0}$ or $Im\varphi _{1}^{0}$ neutral scalar, which is
identified as missing energy, due to the preserved $Z_{4}$ symmetry.
Furthermore, from the lepton Yukawa interactions it follows that the heavy $%
SU(3)_{L}$ singlet exotic charged leptons, i.e. $E_{j}$ ($j=1,2,3$), will
have a dominant decay mode into a SM charged lepton and a neutral CP even $%
\xi _{\rho }$ or CP odd $\zeta _{\rho }$ scalar state, which can also be
identified as missing energy, due to the preserved $Z_{4}$ symmetry. The
exotic $SU(3)_{L}$ singlet up type quarks, i.e. $\widetilde{T}_{1}$ and $%
\widetilde{T}_{2}$ and down type quark, i.e., $B$, are produced in pairs at
the LHC via gluon fusion and the Drell-Yan mechanism, and the charged exotic
leptons $E_{j}$ ($j=1,2,3 $) are also produced in pairs but only via the
Drell-Yan mechanism. Thus, observing an excess of events with respect to the
SM background in the dijet and opposite sign dileptons final states  at the LHC, can be a
signal in support of this model. With respect to the exotic $T$, $%
J^{1}$ or $J^{2}$ quarks, they mainly decay into a top quark and either
neutral or charged scalar. The precise signature of the decays of the
exotic quarks depends on details of the spectrum and other parameters of the
model. The present lower limits on the $Z^{\prime }$ gauge boson mass in $3%
\text{-}3\text{-}1$ models arising from LHC searches, reach around $2.5$
\text{TeV} \cite{Salazar:2015gxa}. These bounds can be translated into
limits of about 6.3 TeV on the $SU(3)_{C}\times SU\left( 3\right) _{L}\times
U\left( 1\right) _{X}$ gauge symmetry breaking scale $v_{\chi }$.
Furthermore, electroweak data from the decays $B_{s,d}\rightarrow \mu
^{+}\mu ^{-}$ and $B_{d}\rightarrow K^{\ast }(K)\mu ^{+}\mu ^{-}$ set lower
bounds on the $Z^{\prime }$ gauge boson mass ranging from $1$ TeV up to $3$
TeV \cite%
{CarcamoHernandez:2005ka,Martinez:2008jj,Buras:2013dea,Buras:2014yna,Buras:2012dp}.
The exotic quarks can be pair produced at the LHC via Drell-Yan and gluon
fusion processes mediated by charged gauge bosons and gluons, respectively.
A detailed study of the exotic quark production at the LHC and the exotic
quark decay modes is beyond the scope of this work and is deferred for a
future publication.

Furthermore, from the quark Yukawa terms of Eq.~(\ref{Lyquarks}), it follows that the flavor changing top
quark decays $t\to hc$, $t\to hu$ and $t\to cZ$ are absent in our model. Besides that, the decays of charged Higgses into a SM up-type and SM down-type quarks, namely, $H^{+}_1\to u_i\bar{d_j}$, $H^{-}_1\to d_i\bar{u_j}$, $H^{+}_2\to u_i\bar{d_j}$, $H^{-}_2\to d_i\bar{u_j}$, ($i,j=1,2,3$) with $H^{\pm}_1=-\rho^{\pm}_1$ and $H^{\pm}_2=-\rho^{\pm}_3$, $\left(u_1,u_2,u_3\right)=\left(u,c,t\right)$ and $\left(d_1,d_2,d_3\right)=\left(d,s,b\right)$, are forbidden at tree level in our model. Out of the charged Higgs decays into SM quarks, only the decays $H^{+}_1\to u_1\bar{d_n}$, $H^{-}_1\to\bar{u_1}d_n$ ($n=1,2$) appear at one loop level whereas the decays $H^{+}_1\to u_2\bar{d_n}$, $H^{-}_1\to\bar{u_2}d_n$ ($n=1,2$) are allowed at two loop level. In addition, the dominant SM leptonic decay modes of the charged Higgses $H^{\pm}_1$ and $H^{\pm}_2$ only appear at one loop level and correspond to the processes $H^{\pm}_1\to \nu_1e^{\pm}$ and $H^{\pm}_2\to \nu_{1}^{c}e^{\pm}$. The remaining decay modes $H^{\pm}_1\to \nu_1\mu^{\pm}$, $H^{\pm}_2\to \nu_{1}^{c}\mu^{\pm}$, $H^{\pm}_1\to \nu_1\tau^{\pm}$, $H^{\pm}_2\to \nu_{1}^{c}\tau^{\pm}$ are very tiny with respect to the decays $H^{\pm}_1\to \nu_1e^{\pm}$ and $H^{\pm}_2\to \nu_{1}^{c}e^{\pm}$, due to the very small mixing angles in the rotation matrix that connects the SM right handed charged leptonic fields in the interaction eigenstates with the physical SM right handed charged leptonic fields.
 Consequently, a measurement of the branching fraction for the $t\to hc$, $t\to hu$, $t\to cZ$, $H^{+}_1\to t\bar{d_j}$, $H^{-}_1\to d_i\bar{t}$, $H^{+}_2\to u_i\bar{d_j}$, $H^{-}_2\to d_i\bar{u_j}$ ($i,j=1,2,3$), $H^{\pm}_1\to \nu_1\mu^{\pm}$, $H^{\pm}_2\to \nu_{1}^{c}\mu^{\pm}$, $H^{\pm}_1\to \nu_1\tau^{\pm}$, $H^{\pm}_2\to \nu_{1}^{c}\tau^{\pm}$ decays at the LHC will be crucial for ruling out this model.

\subsection{Tadpole cancellation mechanisms}
Notice that after $Z_2\times Z_4$ is softly broken, the terms $\overline{E}_{nL}\varphi _{1}^{0}E_{mR}$ and $\left( m_{E}\right) _{nm}%
\overline{E}_{nL}E_{mR}$ $(m,n=2,3)$ will generate a tadpole for $\varphi _{1}^{0}$. Since this contribution is known
to give an infinite value, in order to make the theory renormalizable without giving a VEV to the $\varphi _{1}^{0}$, one has to consider also the contribution to the $\varphi _{1}^{0}$ tadpole arising from the scalar interaction $\omega_{1}\varphi _{1}^{0}(\varphi_{2}^{0})^2$
with the virtual $\varphi_{2}^{0}$ in the loop.
We require that these two tadpoles cancel so that
$\langle\varphi _{1}^{0}\rangle = 0$ be guaranteed at one-loop level. This requirement of tadpole cancellation is
an {\it ad hoc} condition of viability of our model. It implies fine-tuning of the model parameters
$(m_{E})_{nm}$ and $\varphi _{1}^{0}(\varphi_{2}^{0})^2$,
which is unstable under the renormalization flow.
In our model we do not have a symmetry to stabilize the required tadpole cancellation.
Moreover, it is not possible to introduce such a symmetry without \Antonio{a} radical modification of the model structure with all its nice features. The solution to this problem can be expected from the appropriate imbedding of our model into a more fundamental setup with additional symmetries protecting the tadpole cancelation.
Given that this condition relates the parameters of the fermionic and scalar sector one may think of
imbedding
our model into a supersymmetric or warped five-dimensional framework (see Refs. \cite{Ponton:2012bi,Quiros:2013yaa} for recent reviews on extra-dimensions).
Thinking of a supersymmetric (SUSY) version of our model  (for some examples of
SUSY 3-3-1 models see Refs. \cite{Montero:2000ng,CapdequiPeyranere:2001dd,Diaz:2002dx,Rodriguez:2005jt,Sen:2005iw,Sen:2005hj,Dong:2006vk,Sen:2007vx,Dong:2007qc,Huong:2008ww,Huong:2008ia,Huong:2010km,Huong:2012pg,Giang:2012vs,Hue:2013uw,Ferreira:2013nla,Binh:2013axa,Ferreira:2017mvp})
we hope that even in the case of
softly-broken SUSY
the tadpole cancelation would be technically natural. More conservatively we may expect a violation of this cancelation not stronger than logarithms of the high-scale cutoff.
In this case $\langle\varphi _{1}^{0}\rangle \neq 0$, but due to the logarithmic sensitivity to the cutoff, it would be around the electroweak scale. This is phenomenologically safe,
giving rise to tree level mixing $\bar{F}_Lf_R$
between an exotic, $F$, and a SM, $f$, charged fermions. Despite the presence of this mixing terms,
the first and second rows of the up type quark mass matrix as well as the first three rows of the down type quark and charged lepton mass matrices will still be vanishing at tree level, which is a consequence of the symmetries of the model as well as from the fact that the $SU(3)_L$ scalar triplet $\rho$ is inert. This implies that only the top quark and exotic fermions do acquire tree level masses, whereas the remaining SM fermions will be massless at tree level. The masses for the remaining SM fermions will still appear via the radiative seesaw mechanisms described in the previous subsection.
The implementation of supersymmetry or embedding our model in a warped extra-dimensional setup, requires careful studies, which are beyond the scope of the present paper and will be addressed elsewhere.

\section{Quark masses and mixings}

\label{quarksection}

\begin{figure}[tbh]
\resizebox{15.5cm}{20cm} {
\includegraphics{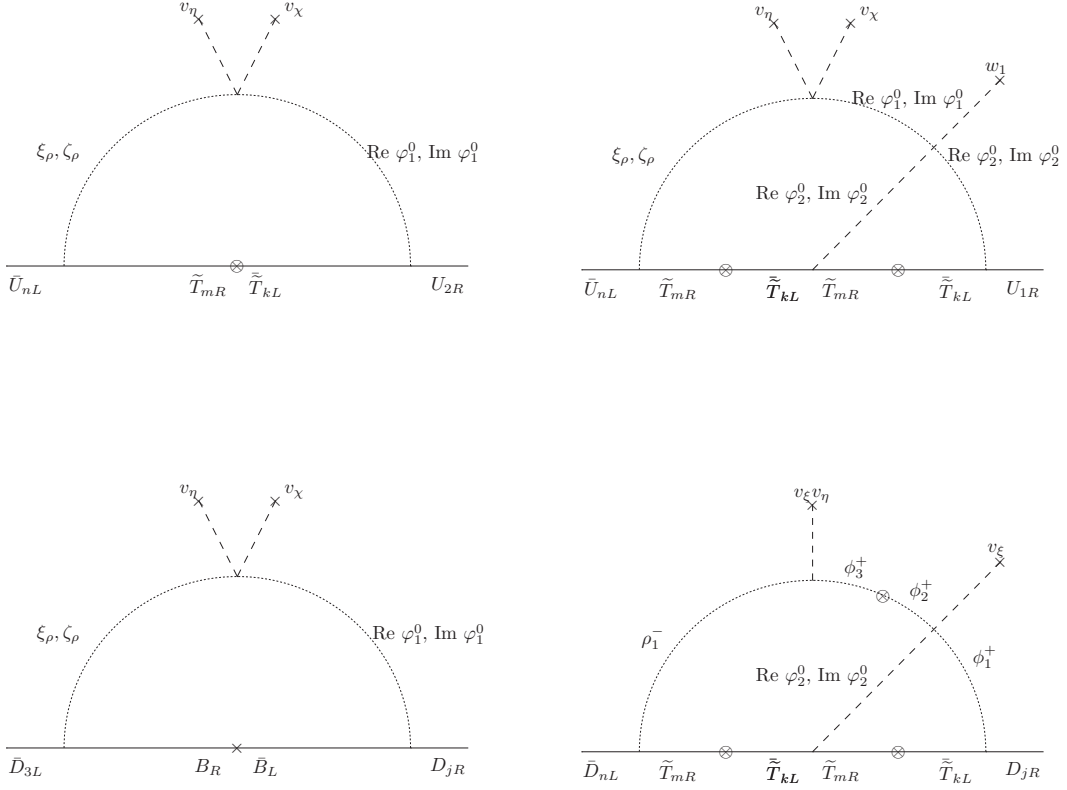}}\vspace{-7cm}
\caption{Loop Feynman diagrams contributing to the entries of the SM quark mass matrices. Here $m,n,k=1,2$ and $j=1,2,3$, whereas $w_1$ corresponds to the mass dimension coefficient of the trilinear scalar coupling $\left(\left(\protect\varphi_{2}^{0}\right)^{\ast }\right) ^{2}\left( \protect\varphi _{1}^{0}\right)^{\ast }$. The cross marks $\times$ and $\otimes$ in the internal lines
correspond to the symmetry preserving and softly breaking mass insertions,
respectively.}
\label{LoopdiagramsUD}
\end{figure}

From the quark
Yukawa interactions (\ref{Lyquarks}) it follows that the SM quark mass matrices
are given by:
\begin{eqnarray}
\label{eq:QMass-matrices}
M_{U}=\left(
\begin{array}{ccc}
\widetilde{\varepsilon }_{11}^{\left( u\right) } & \varepsilon _{12}^{\left(
u\right) } & 0 \\
\widetilde{\varepsilon }_{21}^{\left( u\right) } & \varepsilon _{22}^{\left(
u\right) } & 0 \\
0 & 0 & y%
\end{array}%
\right) \allowbreak \allowbreak \frac{v}{\sqrt{2}},\hspace{1cm}\hspace{1cm}%
M_{D}=\left(
\begin{array}{ccc}
\widetilde{\varepsilon }_{11}^{\left( d\right) } & \widetilde{\varepsilon }%
_{12}^{\left( d\right) } & \widetilde{\varepsilon }_{13}^{\left( d\right) }
\\
\widetilde{\varepsilon }_{21}^{\left( d\right) } & \widetilde{\varepsilon }%
_{22}^{\left( d\right) } & \widetilde{\varepsilon }_{23}^{\left( d\right) }
\\
\varepsilon _{31}^{\left( d\right) } & \varepsilon _{32}^{\left( d\right) }
& \varepsilon _{33}^{\left( d\right) }%
\end{array}%
\right) \allowbreak \allowbreak \frac{v}{\sqrt{2}}
\end{eqnarray}
where $y\simeq 1$ is generated at tree level from the renormalizable Yukawa
interaction $\overline{Q}_{3L}\eta U_{3R}$, thus giving rise to a tree level
top quark mass. Furthermore, $\varepsilon _{n2}^{\left( u\right) }$ ($n=1,2$)
and $\varepsilon _{3j}^{\left( d\right) }$ ($j=1,2,3$) are dimensionless
parameters generated at one loop level, whereas the dimensionless parameters
$\widetilde{\varepsilon }_{n1}^{\left( u\right) }$ and $\widetilde{%
\varepsilon }_{nj}^{\left( d\right) }$ arise at two loop level.
The corresponding Feynman diagrams are shown in Fig.~\ref{LoopdiagramsUD}.

In what follows we will show that the SM quark mass matrices given above are
consistent with the low energy quark flavor data. To this end, and
considering that the $\varepsilon _{n2}^{\left( u\right) }$ ($n=1,2$) and $%
\varepsilon _{3j}^{\left( d\right) }$ ($j=1,2,3$) dimensionless parameters
are generated at one loop level, whereas the dimensionless parameters $%
\widetilde{\varepsilon }_{n1}^{\left( u\right) }$ and $\widetilde{%
\varepsilon }_{nj}^{\left( d\right) }$ arise at two loop level, we choose a
benchmark scenario where we set:
\begin{eqnarray}
\varepsilon _{n2}^{\left( u\right) } &=&a_{n2}^{\left( u\right) }\, l,\hspace{1.5cm}\widetilde{\varepsilon }_{n1}^{\left( u\right)
}=b_{n1}^{\left( u\right) }\,
l^{2},  \notag \\
\varepsilon _{3j}^{\left( d\right) } &=&a_{3j}^{\left( d\right) }\,
l ,\hspace{1.5cm}\widetilde{\varepsilon }_{nj}^{\left( d\right)
}=b_{nj}^{\left( d\right) }\,
l^{2},\hspace{1.5cm}n=1,2,%
\hspace{1.5cm}j=1,2,3,
\label{ep}
\end{eqnarray}%
where $l\approx (1/4\pi)^{2}\approx 2.0 \times \lambda^{4}$ is the loop suppression factor and
$\lambda =0.225$ is the Wolfenstein parameter.
Then we expect in the model that
$a_{n2}^{\left( u\right) }$, $b_{n1}^{\left( u\right) }$, $a_{3j}^{\left(
d\right) }$, $b_{nj}^{\left( d\right) }$ ($n,m=1,2$ and $j=1,2,3$) be
$\mathcal{O}(1)$ parameters.

Let us note that the large amount of independent
model parameters in the fermion and scalar sectors of our model, entering in
the Feynman diagrams contributing to the entries of the SM fermion mass
matrices, can be absorbed in the effective parameters $\varepsilon_{n2}^{\left( u\right) }$,
$\widetilde{\varepsilon }_{n1}^{\left( u\right) }$ $\varepsilon _{3j}^{\left( d\right) }$,
$\widetilde{\varepsilon }_{nj}^{\left( d\right) }$ ($n,m=1,2$ and $j=1,2,3$) given
by Eq. (\ref{ep}).
They amount to 26 real free parameters, which is a large number compared with
the number of quark sector observables with the experimental values
\footnote{We use the experimental values of the quark masses at the $M_{Z}$ scale, from
Ref. \cite{Bora:2012tx}, which are similar to those in \cite{Xing:2007fb}.
The experimental values of the CKM parameters are taken from Ref. \cite{Olive:2016xmw}.
}
\begin{eqnarray}
\label{eq:Qsector-observables}
&&m_{u}(MeV) = 1.45_{-0.45}^{+0.56},\hspace{3mm}
m_{d}(MeV) = 2.9_{-0.4}^{+0.5}, \hspace{3mm}
m_{s}(MeV) = 57.7_{-15.7}^{+16.8}, \\
\nonumber
&&m_{c}(MeV) = 635\pm 86, \hspace{3mm}
m_{t}(GeV) =172.1\pm 0.6\pm 0.9, \hspace{3mm}
m_{b}(GeV) = 2.82_{-0.04}^{+0.09}, \hspace{3mm}\\
\nonumber
&&\sin \theta _{12} = 0.2254,\hspace{3mm}
\sin \theta _{23} = 0.0414, \hspace{3mm}
\sin \theta _{13} = 0.00355, \hspace{3mm}
J = 2.96_{-0.16}^{+0.20}\times 10^{-5}.
\end{eqnarray}
being $t, u, c, d, s, b$ quark masses,
$\theta_{12}$, $\theta_{23}$, $\theta_{13}$ mixing angles and the Jarlskog parameter.
Therefore, the model in its present form does not predict these observables.
However, as we already commented, we only pretend to reproduce the hierarchy of the quark
masses via the loop suppression predicted by the model and expressed by Eq.~(\ref{ep}).
To wit, we consider the mass matrices (\ref{eq:QMass-matrices}) with the hierarchical matrix elements (\ref{ep}) predicted by the model.  For these matrices we
look for the eigenvalue problem solutions reproducing the values in
Eq.~(\ref{eq:Qsector-observables}), under the condition that
$a^{(u,d)}, b^{(u,d)}$  be most close to $\mathcal{O}(1)$. Applying the standard procedure we find a solution
\begin{eqnarray}
\label{eq:fit-Q}
a_{12}^{\left( u\right) }&\simeq& a_{22}^{\left( u\right) } \approx 0.5, \hspace{3mm}
b_{11}^{\left( u\right)} \approx 0.25, \hspace{3mm} b_{21}^{\left(u\right)} \approx 0.7,\\
\nonumber
a_{31}^{\left( d\right) } &\simeq &-1.6,\hspace{3mm}
a_{32}^{\left(d\right) }\simeq -2.2, \hspace{3mm} a_{33}^{\left( d\right) }\simeq 1.6,\\
\nonumber
b_{11}^{\left( d\right) }&\simeq& -12.3 - 0.7 i,
\hspace{3mm}b_{12}^{\left( d\right) }\simeq-7.6 - 1.0 i,\hspace{3mm}
b_{13}^{\left( d\right) }\simeq 11.6 + 0.7 i,  \notag \\
\nonumber
b_{21}^{\left( d\right) } &\simeq &-14.3 + 0.7 i,\hspace{3mm}
b_{22}^{\left(d\right) }\simeq -5.6 + 1.0 i,\hspace{3mm}b_{23}^{\left( d\right) }\simeq
14.8 - 0.7 i.
\end{eqnarray}
The above values reproduce exactly the central values in (\ref{eq:Qsector-observables}).
The absolute values of the parameters in the first two rows are $\mathcal{O}(1)$. The values of the  remaining two-loop level parameters are around $\sim$10, but this is still within the ballpark of the same loop level, since the loop-suppression factor is $l \sim 10^{-2}$ (see Eq.~(\ref{ep})).
Thus the model reproduces fairly well the hierarchical structure of the observable quark mass spectrum as a result of the sequential loop suppression  mechanism, where the top quark mass is generated at tree level, the masses for the bottom and charm quarks arise at one loop level and the light up, down and strange quarks get their masses at two loop level.

\section{Lepton masses and mixings}

\label{leptonmassesasndmixings}
\begin{figure}[tbh]
\begin{center}
\resizebox{15.5cm}{20cm}{
\includegraphics[scale=0.8]{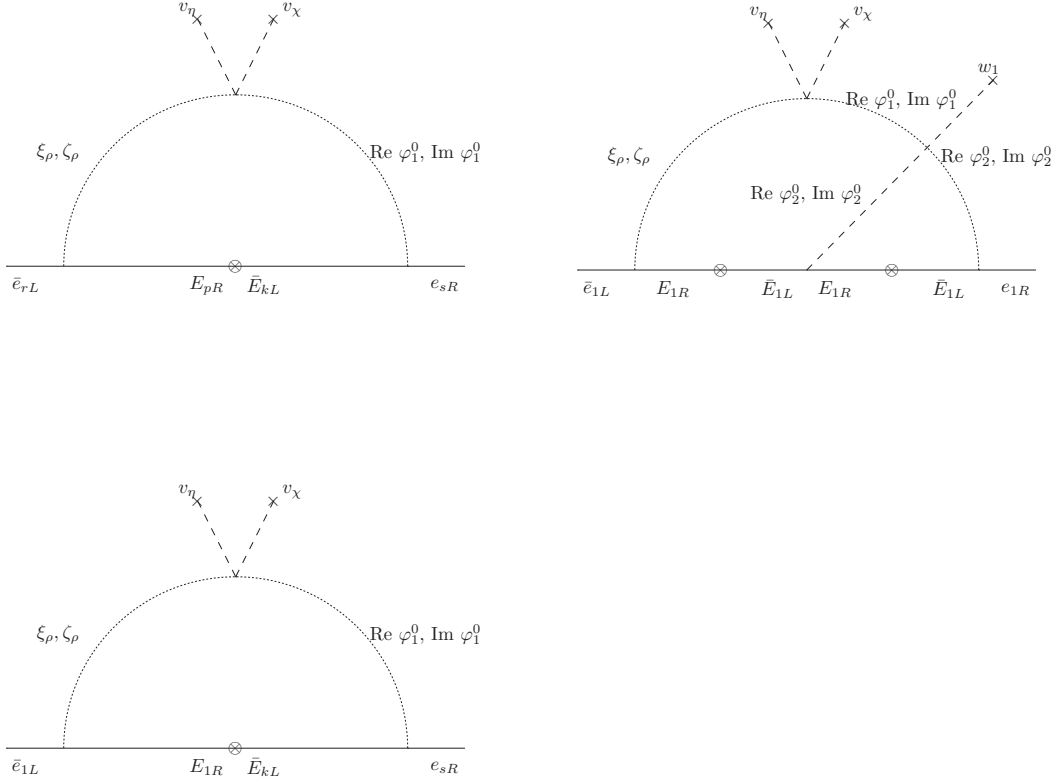}}
\end{center}
\par
\vspace{-9cm} 
\caption{Loop Feynman diagrams contributing to the entries of the charged
lepton mass matrices. Here $k,p,r,s=2,3$ and $w_1$ corresponds to
the mass dimension coefficients of the trilinear scalar coupling $\left(\left(%
\protect\varphi_{2}^{0}\right) ^{\ast }\right) ^{2}\left( \protect\varphi %
_{1}^{0}\right)^{\ast }$.
The cross marks in the internal lines denote the $%
m_{E}$ mass insertions from Eq.~(\protect\ref{eq:Fermion-Soft-terms}). }
\label{Loopdiagrams3}
\end{figure}

The charged lepton masses are generated by the charged lepton Yukawa terms in Eq.~(\ref{Lyleptons}) via the loop diagrams shown in Fig~\ref{Loopdiagrams3}. The corresponding charged lepton mass matrix takes the form:
\begin{equation}
M_{l}=\left(
\begin{array}{ccc}
\widetilde{\varepsilon }_{11}^{\left( l\right) } & \varepsilon _{12}^{\left(
l\right) } & \varepsilon _{13}^{\left( l\right) } \\
0 & \varepsilon _{22}^{\left( l\right) } & \varepsilon _{23}^{\left(
l\right) } \\
0 & \varepsilon _{32}^{\left( l\right) } & \varepsilon _{33}^{\left(
l\right) }%
\end{array}%
\right) \frac{v}{\sqrt{2}},  \label{Ml}
\end{equation}%
where $\varepsilon _{jn}^{\left( l\right) }$ ($n=2,3$ and $j=1,2,3$) are
dimensionless parameters generated at one loop level, whereas the
dimensionless parameter $\widetilde{\varepsilon }_{11}^{\left( l\right) }$
arises at two loop level.
In order to express the loop order suppression explicitly we define
new parameters
\begin{equation}
\label{eq:CL-hierarchy}
\varepsilon _{j3}^{\left( l\right) }=a_{j3}^{\left( l\right) }\cdot l,%
\hspace{1.5cm}\varepsilon _{j2}^{\left( l\right) }=a_{j2}^{\left( l\right)
}\cdot l,\hspace{1.5cm}\widetilde{\varepsilon }_{11}^{\left( l\right)
}=a_{11}^{\left( l\right) }\cdot l^{2},\hspace{1.5cm}j=1,2,3.
\end{equation}
Here $l=(1/4\pi)^{2}\approx 2.0\times \lambda^{4}$  is the loop suppression factor introduced after Eq.~(\ref{ep}). Having 7 complex parameters, the model does not pretend to predict the charged lepton masses, but only reproduce the observed mass hierarchy.   Therefore, again, as in the quark sector, we are looking for values of the $a^{(l)}$ parameters so that on one hand they reproduce the observable central values of the charged lepton masses
$m_{e}=0.487$MeV, $m_{\mu} = 102.8$MeV, $m_{\tau} = 1.75$GeV and on the other hand
the condition $|a^{(l)}|\sim \mathcal{O}(1)$ is  achieved as close as possible.
A benchmark point in the model parameter space of this kind is
\begin{eqnarray}\label{eq:CL-BMP-1}
&&a^{(l)}_{11} \approx a^{(l)}_{22} \approx  -0.2, \hspace{5mm}
a^{(l)}_{12} \approx a^{(l)}_{32} \approx -0.14, \hspace{3mm}
a^{(l)}_{13} \approx a^{(l)}_{23} \approx a^{(l)}_{33} \approx 1.1.
\end{eqnarray}
All the values are not unnaturally small compared to the loop hierarchy
(2-loop level)/(1-loop level)$\sim l \approx 6.3 \times 10^{-3}$.   Thus, as in the quark sector, the model proves to be able to reproduce the observed charged lepton mass hierarchy by a sequential loop suppression.

\begin{figure}[tbh]
\vspace{-1.5cm}
\resizebox{15.5cm}{12.4cm}{\includegraphics[scale=0.4]{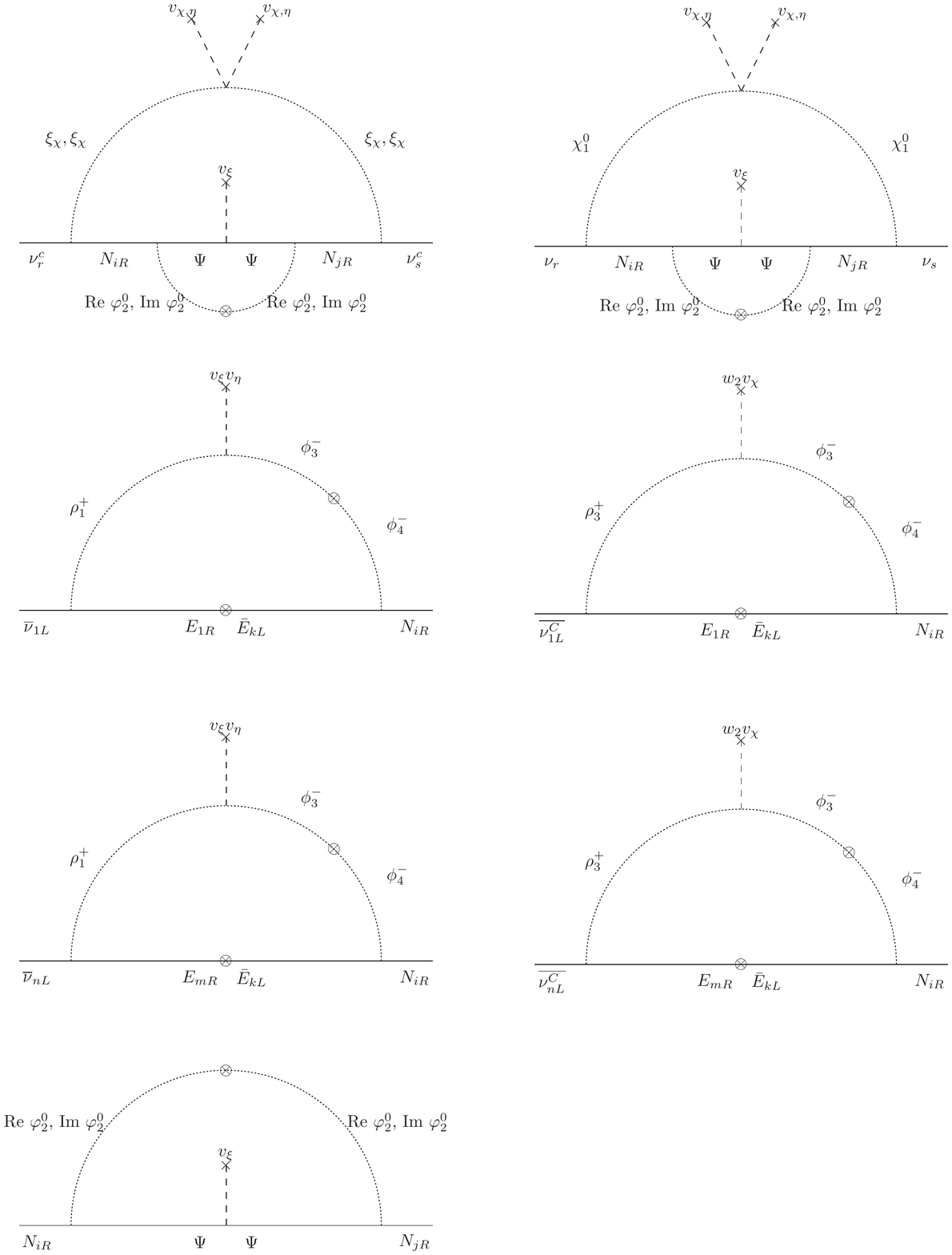}}\vspace{-0.6cm}
\resizebox{15.5cm}{12.4cm}{\includegraphics[scale=0.4]{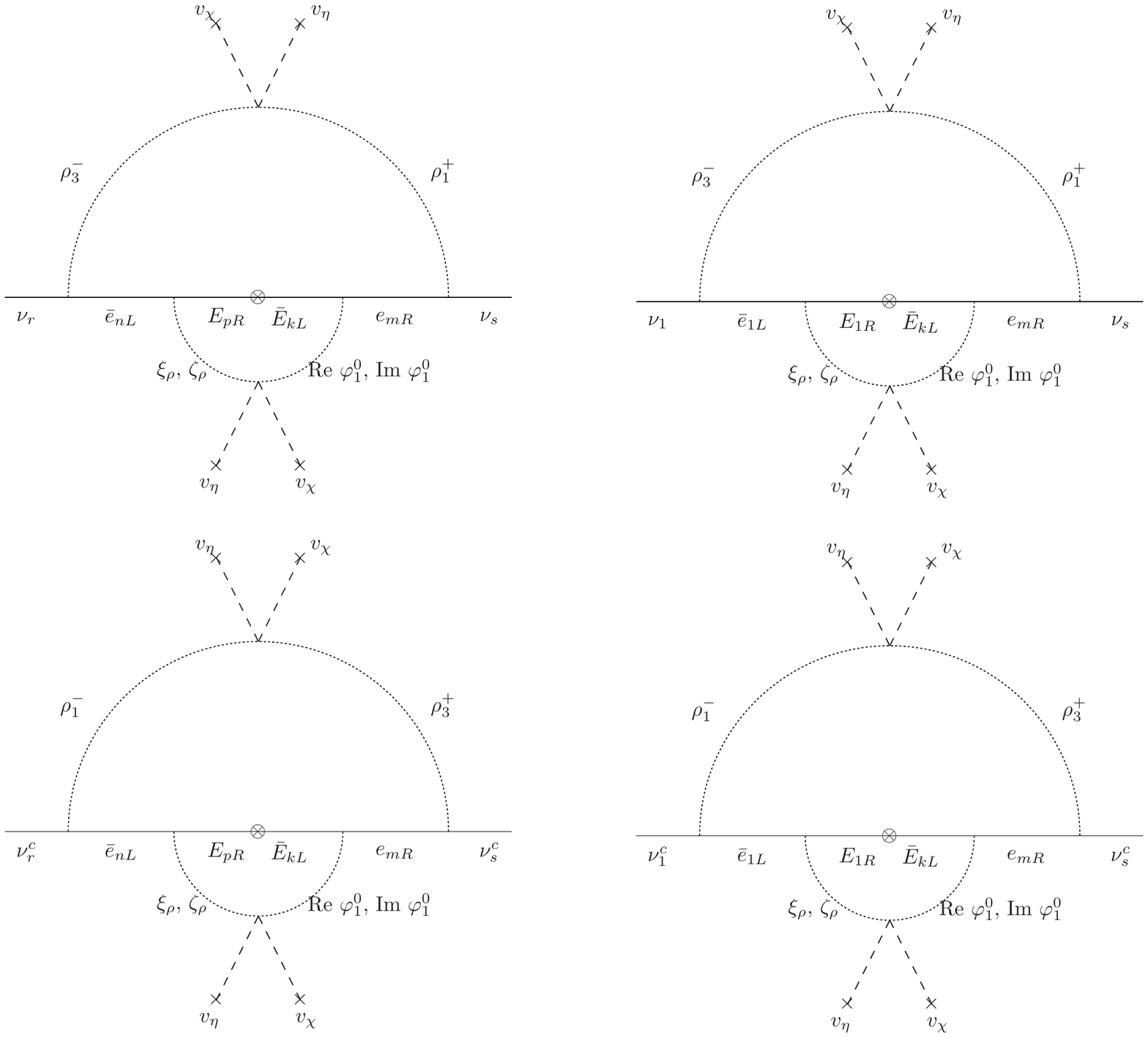}}\vspace{-4.2cm}
\caption{Loop Feynman diagrams contributing to the entries of the neutrino
mass matrix. Here $n,m,k,p,r,s=2,3$ and $i,j=1,2,3$, whereas $w_{2}$ corresponds to the mass dimension coefficient of the trilinear
scalar coupling $\protect\phi _{3}^{+}\protect\rho^{\dagger}\protect\chi$. The cross mark  $\otimes$ in the internal lines correspond to
the softly breaking mass insertions.
}
\label{Loopdiagrams2}
\end{figure}

From the neutral lepton Yukawa interactions in Eq.~(\ref{Lyleptons}) we find the neutral lepton mass terms:
\begin{equation}
-L_{g mass}^{\left( \nu \right) }=\frac{1}{2}\left(
\begin{array}{ccc}
\overline{\nu _{L}^{C}} & \overline{\nu _{R}} & \overline{N_{R}}%
\end{array}%
\right) M_{\nu }\left(
\begin{array}{c}
\nu _{L} \\
\nu _{R}^{C} \\
N_{R}^{C}%
\end{array}%
\right) +m_{\Psi }\overline{\Psi _{R}^{c}}\Psi _{R}+h.c,  \label{Lnu}
\end{equation}%
where the neutrino mass matrix $M_{\nu }$
\begin{equation}
M_{\nu }=\left(
\begin{array}{ccc}
M_{1} & 0_{3\times 3} & M_{3} \\
0_{3\times 3} & M_{2} & M_{4} \\
M_{3} & M_{4} & \mathcal{M}
\end{array}%
\right) ,  \label{M3}
\end{equation}%
is generated by the loop diagrams shown in Fig.~\ref{Loopdiagrams2}.
The sub-matrices $M_{1}$, $M_{2}$, $M_{3}$, $M_{4}$ and $\mathcal{M} $ are
given by
\begin{eqnarray}
M_{1} &=&\left(
\begin{array}{ccc}
0 & a_{12} & a_{13} \\
a_{12} & a_{22} & a_{23} \\
a_{13} & a_{23} & a_{33}%
\end{array}%
\right),\hspace{1cm}M_{2}=\left(
\begin{array}{ccc}
0 & b_{12} & b_{13} \\
b_{12} & b_{22} & b_{23} \\
b_{13} & b_{23} & b_{33}%
\end{array}%
\right),  \\
M_{3} &=& \left(
\begin{array}{ccc}
\epsilon_{11} & \epsilon_{12} & \epsilon_{13} \\
\epsilon_{21} & \epsilon_{22} & \epsilon_{23} \\
\epsilon_{31} & \epsilon_{32} & \epsilon_{33}
\end{array}%
\right) \frac{v}{\sqrt{2}},\hspace{1cm}M_{4}=\left(
\begin{array}{ccc}
\varepsilon _{1} & \varepsilon _{2} & \varepsilon _{3} \\
d_{1} & d_{2} & d_{3} \\
d_{4} & d_{5} & d_{6}%
\end{array}%
\right) \frac{v_{\chi }}{\sqrt{2}},\hspace{1cm}\mathcal{M} =\left(
\begin{array}{ccc}
\mathcal{M} _{11} & \mathcal{M} _{12} & \mathcal{M} _{13} \\
\mathcal{M} _{12} & \mathcal{M} _{22} & \mathcal{M} _{23} \\
\mathcal{M} _{13} & \mathcal{M} _{23} & \mathcal{M} _{33}%
\end{array}%
\right), \notag \label{M4}
\end{eqnarray}
where
the matrix elements $d_{l}$ ($l=1,2,\cdots ,6$) arise at
tree level, $\epsilon _{ij}$, $\varepsilon _{j}$ and $\mathcal{M} _{ij}$  ($i,j=1,2,3$) at one loop level, whereas $a_{nm}$, $a_{1n}$, $b_{1n}$ and $b_{nm}$ ($n,m=2,3$) arise at two loop level.
Let us note that $a_{1n}$ and $b_{1n}$ are generated by the
$h_{\rho 11}^{\left( L\right) }$
and
$h_{\rho nm}^{\left( L\right) }$
terms in Eq.~(\ref{Lyleptons}). These terms give rise to the four Feynman diagrams shown in the last two lines of Fig.~\ref{Loopdiagrams2}.
Consequently, the light active
neutrino masses are generated by a combination of linear and inverse seesaw
mechanisms at two loop level.

By performing the perturbative block diagonalization of the $9\times 9$
neutrino mass matrix $M_{\nu }$ of Eq.~(\ref{M3}), which is shown in Appendix \ref{DiagonalizationMnu}, we find that the light
active neutrino mass matrix has the form:
\begin{eqnarray}
M_{1\nu } &=&M_{1}+\frac{1}{16}M_{3}\left( M_{4}\right)
^{-2}M_{3}^{T}M_{3}\left( M_{4}\right) ^{-1}M_{3}^{T}M_{3}\left(
M_{4}\right) ^{-2}M_{3}^{T}  \\
&&+\frac{1}{8}M_{3}\left( M_{4}\right) ^{-1}\left( \mathcal{M} -M_{3}^{T}M_{3}\left(
M_{4}\right) ^{-1}\right) \left( M_{4}\right) ^{-1}\left( \mathcal{M} -\left(
M_{4}\right) ^{-1}M_{3}^{T}M_{3}\right) \left( M_{4}\right) ^{-1}M_{3}^{T}\notag
\label{M1nu}
\end{eqnarray}%
whereas the sterile neutrino mass matrices are given by:
\begin{eqnarray}
M_{2\nu } &=&-M_{4}  \notag \\
M_{3\nu } &=&M_{4}+\frac{1}{\sqrt{2}}\left( M_{3}^{T}M_{3}\left(
M_{4}\right) ^{-1}+\left( M_{4}\right) ^{-1}M_{3}^{T}M_{3}\right) .
\label{Sterile}
\end{eqnarray}
Let us analyze Eq.~(\ref{M1nu}) and see what are the typical mass scales of the model, which allow us to reproduce the neutrino mass scale of $m_{\nu}\sim$50 meV.
The non-zero matrix elements of $M_{1}$ are determined by the 2-loop diagrams in Fig.~\ref{Loopdiagrams2}.
For the benchmark region where $m_{\Psi} = y_{\Psi} v_{\xi}  \gg m_{\phi^{0}_{2}}, \mu_{1}$ their contribution is
\begin{eqnarray}\label{eq:Loop-nuM12}
a_{ij} &\sim& \alpha_{1} \left(\frac{1}{4\pi}\right)^{2}\left(\frac{v_{\chi}}{v}\right)^{2} \frac{\mu^{2}_{1}}{m_{\Psi}} \log\left(m_{\Psi}/m_{\phi^{0}_{2}} \right).
\end{eqnarray}
This 2-loop-level contribution has typical inverse or linear seesaw structure proportional to the soft symmetry breaking parameter $\mu^{2}_{1}$. This parameter is stable against radiative corrections due to the model symmetries and, therefore, any of its possible values is technically natural. Then we can choose it arbitrarily small to adjust the observable neutrino mass scale $m_{\nu}\sim$ 50 meV.
Note that $a_{ij}\rightarrow 0$ in the limit $m_{\Psi}\rightarrow 0$, since $m_{\Psi}$ is the only
Lepton Number Violating parameter in our model.

The second term in Eq.~(\ref{M1nu})  is of the order
\begin{eqnarray}\label{eq:Tree-level-Nu-1}
\left(\mbox{2nd term in Eq.~(\ref{M1nu})} \right) &\sim& \left(\frac{v}{v_{\chi}}\right)^{5}v.
\end{eqnarray}
From the condition $\left(M_{1\nu}\right)_{ij}\lesssim m_{\nu}\sim 50$ meV $(i,j=,1,2,3)$, we find that the scale of the first stage of symmetry breaking (\ref{Group}) is limited to
\begin{eqnarray}\label{eq:lim-v-chi}
&&v_{\chi}\gtrsim 90\, \mbox{TeV}
\end{eqnarray}

It is worth mentioning that, as follows from Eqs.~(\ref{M1nu}) and (\ref{Sterile}), the physical neutrino eigenstates include three
active neutrinos and six exotic neutrinos.
Due to the structure of $M_{3,4}$ in Eq.~(\ref{M4}) and using Eq. (\ref{Sterile}), it is shown in Appendix \ref{Sterilespectrum} that the second and third generation of exotic neutrinos arising from $M_{2\nu}$ and $M_{3\nu }$ have $\mathcal{O}(10)\mbox{TeV}$ scale masses, whereas the first generation ones from $M_{2\nu}$ and $M_{3\nu }$ have masses at the electroweak symmetry breaking scale. The $\mathcal{O}(10)\mbox {TeV}$ scale exotic neutrinos of $M_{2\nu}$ have a small
splitting of $\sim \frac{v^{2}}{v_{\chi }}$ with respect to the ones of $M_{3\nu }$, as indicated by Eq. (\ref{Sterile}).
These heavy quasi Dirac neutrinos can be produced in
pairs at the LHC, via a Drell-Yan mechanism, mediated by a heavy non Standard
Model neutral gauge boson $Z^{\prime }$. The heavy quasi Dirac neutrinos can
decay into a Standard Model charged lepton and $W$ gauge boson, due to their
mixings with the light active neutrinos. Thus, the observation of an excess
of events in the dilepton final states with respect to the SM background at the LHC
would be a signal supporting this model. Studies of inverse seesaw neutrino signatures at the LHC and ILC as well as the production of Heavy neutrinos at the LHC are performed in Refs. \cite{Das:2012ze,Das:2016hof}. A detailed study of the
collider phenomenology of this model is beyond the scope of the present
paper and is left for future studies.

\section{Discussions and conclusions}

\label{conclusions}

We have built the first renormalizable extension of the 3-3-1 model with
$\beta =-\frac{1}{\sqrt{3}}$, which explains the SM fermion mass hierarchy by
a sequential loop suppression. Our model, based on the 3-3-1 symmetry extended
with the $U(1)_{L_g}\times Z_{4}\times Z_{2}$ group
is consistent
with the low energy fermion flavor data. In the model only the top quark and
the charged exotic fermions acquire tree level masses, whereas the remaining
SM fermions get their masses via radiative corrections: 1 loop bottom,
charm, tau and muon masses; 2-loop masses for the light up, down, strange
quarks as well as for the electron. Furthermore, the light active neutrinos
acquire their masses from a combination of linear and inverse seesaw
mechanisms at two loop level. In our model the quark and lepton mixings
arise from radiative effects. At tree level there is no quark mixing, the
mixing angles in the quark sector are generated from a combination of one
and two loop level radiative seesaw mechanisms. In the lepton sector, the
contribution to the leptonic mixing angles coming from the charged leptons
arise at one loop level, whereas the mixings in the light active neutrino
sector are generated from a two loop level radiative seesaw mechanism.

Furthermore our model predicts the absence of the decays $t\to hc$, $t\to hu$, $t\to cZ$, $H^{+}_1\to t\bar{d_j}$, $H^{-}_1\to d_i\bar{t}$, $H^{+}_2\to u_i\bar{d_j}$, $H^{-}_2\to d_i\bar{u_j}$ ($i,j=1,2,3$), $H^{\pm}_1\to \nu_1\mu^{\pm}$, $H^{\pm}_2\to \nu_{1}^{c}\mu^{\pm}$, $H^{\pm}_1\to \nu_1\tau^{\pm}$, $H^{\pm}_2\to \nu_{1}^{c}\tau^{\pm}$, which implies that a measurement of the branching fraction for these decays at the LHC will be crucial for ruling out the model. Consequently, charged Higgses can be searched at the LHC through the their decay into a SM up-type (down-type) and a exotic SM down-type $B$ ($\widetilde{T}_{k}$ ($k=1,2$) up-type) quarks, as well as into a exotic charged lepton and neutrino. Since
the heavy exotic $SU(3)_{L}$ singlet down (up) type quark(s), i.e., $B$ ($\widetilde{T}_{n}$ ($n=1,2$))\
will decay predominantly into a SM down (up) type quark and the $Re%
\varphi _{1}^{0}$ or $Im\varphi _{1}^{0}$ neutral scalar (which is
identified as missing energy, due to the preserved $Z_{4}$ symmetry), it follows that the observation of an excess of events with respect to the
SM background in the dijet final states at the LHC can be a signal of charged Higgs decays of this model. Finally, it is worth mentioning that since charged exotic fermions are produced in pairs, and they predominantly decay into a SM charged fermion and a electrically neutral $Z_4$ charged scalar (identified as a missing energy), observing an excess of events with respect to the
SM background in the dijet and opposite sign dilepton final states at the LHC can be a
signal in support of this model.
A detailed study of the exotic charged fermion production at the LHC and the exotic
charged fermion decay modes is beyond the scope of this work and is deferred for a
future publication.

The final remark deals with the possible DM candidates in our model,
which could be either the right handed Majorana neutrinos $N_{iR}$ ($i=1,2,3$), $\Psi_R$, or the lightest scalars $\varphi^R_{1} \equiv$ Re$\varphi^0_{1}$,
$\varphi^I_{1} \equiv$ Im$\varphi^0_{1}$ as well as $\varphi^R_{2} \equiv$ Re$\varphi^0_{2}$,
$\varphi^I_{2} \equiv$ Im$\varphi^0_{2}$.
Let us note that the masses $m_{1}^{R,I}, m_{2}^{R}$ of the scalars $\varphi^{R,I}_{1}$,
$\varphi^{R}_{2}$ and the fermion
$\Psi_{R}$ mass $m_{\Psi}$ are arbitrary parameters, since the corresponding mass terms are compatible with all the symmetries of the model, while $\varphi^{I}_{2}$ squared mass is
$(m^{R}_{2})^{2} - 4 \mu_{1}^{2}$. The mass splitting parameter $\mu_{1}^{2}$ is the soft $Z_{4}$ breaking mass (\ref{eq:Scalar-Soft-terms}), which already showed up in the light neutrino sector (\ref{eq:Loop-nuM12}). Since the light neutrino mass scale should be small, then the
$\varphi^{R}_{2}-\varphi^{I}_{2}$ mass splitting should not be very large. A superficial survey
shows that $N_{iR}$ could be a DM candidate only in a rather restricted domain of the model parameter space, due to the presence of $N_{R}-\nu_{L}$-mixing $U_{\nu N}$ at least at one-loop level (fifth diagram in Fig.~\ref{Loopdiagrams2}). As a result, there is the SM charged current decay
$N_{R}\rightarrow e^{-}_{L} \nu_{1} e^{+}_{R}$.  The requirement that the DM lifetime be greater than the universe lifetime sets stringent constraints on $U_{\nu N}$. We do not analyze the impact of this constraint on the model parameter space and the possible correlations of the DM and the light neutrino sectors. Instead we consider the other more viable DM candidates.  The gauge group singlet $\Psi_{R}$ is one of them, if its mass satisfies the condition $m_{\Psi} < m^{R}_{2}$, and then, as follows from Eq.~(\ref{Lyleptons}), it does not decay at tree level.
Assuming that our model be valid only up to some high-energy scale
$\Lambda\gg v_{\chi}$,  we have to consider the possible non-renormalizable operators induced by the physics beyond this scale. It is easy to check that all such operators compatible with the symmetry group $\cal{G}$
of our model
involve the exotic scalar $\varphi^{0}_{2}$.  The lowest dimensional operator  is
\begin{eqnarray}\label{eq:eff-operat-Psi}
&&\frac{1}{\Lambda^{3}}\left(\bar{L} \Psi_{R}\right) \left( \overline{e}_{R} L\right)
\varphi^{0}_{2}
\end{eqnarray}
Therefore, with the condition $m_{\Psi} < m^{\Psi}_{2}$ the non-renormalizable operators do not lead to kinematically allowed decays of $\Phi_{R}$, making it a stable DM particle.
There is also a viable scalar DM candidate $\varphi^{0}$ in our model. This is the lightest of the exotic scalars
Re$\varphi^0_{1,2}$, Im$\varphi^0_{1,2}$, which is also lighter than the exotic charged fermions, as well as lighter than $\Psi_{R}$, and then, as follows from Eqs.~(\ref{Lyquarks}),  (\ref{Lyleptons}),  its tree-level decays are kinematically forbidden.
However, as before, we check the possible non-renormalizable operators originating from the scales
$\Lambda$, above the theoretical validity of our model. In the case of the DM candidate
$\varphi^{0} \equiv$ Re$\varphi^{0}_{1}$ or  Im$\varphi^{0}_{1}$, we find the dominant operator
\begin{eqnarray}\label{eq:Eff-Operator-Phi-1}
&&\frac{1}{\Lambda^{2}}\epsilon_{abc}\left(\eta^{\dagger}\right)^a\left(\chi^{\dagger}\right)^b\varphi^{0}_{1}\overline{L}^c_{1}e_{kR}\ \ \ \ \ \mbox{for}\ \ \ \ \ k=2,3
\end{eqnarray}
compatible with all the symmetries of our model.
This operator induces the decays $\varphi^{0}\rightarrow e^{+}_{1}e^{-}_{2,3}\xi_{\eta}$, $\varphi^{0}\rightarrow e^{+}_{1}e^{-}_{2,3}\zeta_{\eta}$, $\varphi^{0}\rightarrow e^{+}_{1}e^{-}_{2,3}\xi_{\chi}$, $\varphi^{0}\rightarrow e^{+}_{1}e^{-}_{2,3}\zeta_{\chi}$, $\varphi^{0}\rightarrow e^{+}_{1}e^{-}_{2,3}$, respectively. Here
$\xi_{\eta}=\cos\alpha h^{0}+\sin\alpha H^{0}_1$, $\zeta_{\eta}\simeq G^{0}_1$, with $\tan\theta\sim \mathcal{O}(\frac{v}{v_ {\chi}})$ and $\alpha
$ a mixing angle, which depends on the scalar potential parameters. Furthermore, $h$ is the $126$ GeV SM Higgs boson, $H^{0}_1$ is one of the physical heavy neutral Higges, whereas $G^{0}_1$ is the Goldstone boson associated with the longitudinal component of the $Z$ gauge boson.
For the scenario where the scalar $\varphi^{0}$ is heavier than the $126$ GeV Higgs, the partial decay rates of the kinematically allowed processes
can be estimated as
\begin{eqnarray}
\label{eq:DecRate}
&&\Gamma (\varphi^{0}\rightarrow Ze^{+}_{1}e^{-}_{2,3})\simeq \Gamma (\varphi^{0}\rightarrow\zeta_{\eta}e^{+}_{1}e^{-}_{2,3})\sim \Gamma (\varphi^{0}\rightarrow he^{+}_{1}e^{-}_{2,3})\sim m^3_{\varphi^{0}}\frac{v^2_{\chi}}{\Lambda^{4}},\nonumber\\
&&\Gamma (\varphi^{0} \rightarrow e^{+}_{1}e^{-}_{2,3}) \sim m_{\varphi^{0}}\left(\frac{v_{\chi}v_{\eta}}{\Lambda^2}\right)^{2}.
\end{eqnarray}
Requiring that the DM candidate $\varphi^{0}$ lifetime be greater than the universe lifetime
$\tau_{u} \approx 13.8$ Gyr, taking into account the limit (\ref{eq:lim-v-chi}) and assuming $m_{\varphi^{0}} \sim 1$ TeV, we estimate the cutoff scale of our model
\begin{eqnarray}\label{eq:CutOff}
&& \Lambda > 3\times 10^{10} \mbox{GeV}.
\end{eqnarray}
Thus we conclude that under the above specified conditions the model contains viable fermionic
$\Psi_{R}$ and scalar $\varphi^{0}$ DM candidates.
A detailed study of the dark matter constraints in our model is beyond the scope of the present paper and will be considered elsewhere.

\section*{Acknowledgments}

This research has received funding from Fondecyt (Chile) grants No.~1170803,
No.~1150792, No.~1180232, No.~3150472 and by CONICYT (Chile) Ring ACT1406,
PIA/Basal FB0821, the UTFSM grant FI4061, the Vietnam National Foundation
for Science and Technology Development (NAFOSTED) under Grant No.
103.01-2017.356.  A.E.C.H is very grateful to the Institute of Physics,
Vietnam Academy of Science and Technology for the warm hospitality and for
fully financing his visit.

\appendix

\section{Generalized lepton number}
\label{gln}

Since the lepton and anti-lepton lie in the triplet,  the lepton number operator $L$  does not commute with the $SU(3)_C\times SU(3)_L\times U(1)_X$ gauge symmetry and has the form
\cite{Chang:2006aa}
\begin{equation}
L=\frac{4}{\sqrt{3}}T_{8}+L_g=\left(
\begin{array}{ccc}
\pm\frac{2}{3}+L_g & 0 & 0 \\
0 & \pm\frac{2}{3}+L_g & 0 \\
0 & 0 & \mp\frac{4}{3}+L_g
\end{array}%
\right),\label{Leptonnumber}
\end{equation}%
where the upper and lower signs correspond to triplet and antitriplet of $SU(3)_L$, respectively.
Here
$L_g$ is a conserved charge corresponding to the $U(1)_{L_g}$ global symmetry, which commutes with the gauge symmetry.
According to the analysis done in Ref. \cite{Chang:2006aa}, the $SU(3)_L$ Higgs triplets $\chi$ and $\eta$ have different $L_g$ charges, which are given by:
\be L_g(\chi) = \frac 4 3 \, ,\hspace{1cm}  \,  L_g(\eta) = -\frac 2 3 .\label{gtl}
\ee
From the application of (\ref{Leptonnumber}) to (\ref{scalarsector1}), it follows that
the top component of $\chi$ and
the bottom one of $\eta$ carry lepton number $L(\chi_1^0) = - L(\eta_3^0) = 2$
whereas the other components
do not $L(\chi_3^0) = L(\eta_1^0) = 0 $.
\section{Perturbative diagonalization of the neutrino mass matrix}
\label{DiagonalizationMnu}
In this appendix we show explicitly the perturbative diagonalization of the $9\times 9$ neutrino mass matrix $M_{\nu}$  of our model, which is given by Eqs.~(\ref{Lnu})-(\ref{M4}).
The elements of the submatrices $M_{1,2,3,4}$ obey the following hierarchy:
\begin{equation}
\left( M_{1}\right) _{ij}\sim \left( M_{2}\right) _{ij}\ll \mathcal{M} _{ij}\ll \left(
M_{3}\right) _{ij}\ll \left( M_{4}\right) _{ij}
\label{Hierarchymatrices}
\end{equation}
with $i,j=1,2,3$.

We first apply the following orthogonal transformation to the matrix $M_{\nu}$:
\begin{eqnarray}
S^{T}_{\nu}M_{\nu}S_{\nu}&\simeq &\left(
\begin{array}{ccc}
M_{1} & \frac{1}{\sqrt{2}}M_{3} & \frac{1}{\sqrt{2}}M_{3} \\
\frac{1}{\sqrt{2}}M_{3}^{T} & M_{4} & \frac{1}{2}\mathcal{M} \\
\frac{1}{\sqrt{2}}M_{3}^{T} & \frac{1}{2}\mathcal{M} & -M_{4}%
\end{array}%
\right),\hspace{1cm}S_{\nu}=\left(\begin{array}{ccc}
I & 0 & 0 \\
0 & \frac{1}{\sqrt{2}}I & \frac{1}{\sqrt{2}}I \\
0 & -\frac{1}{\sqrt{2}}I & \frac{1}{\sqrt{2}}I
\end{array}\right)
\label{MS}
\end{eqnarray}
where $I$ is the $3\times 3$ identity matrix.

Then, a second orthogonal transformation is applied under the matrix $M_{\nu}$, as follows:
\begin{eqnarray}
&&R^{T}_{1\nu}S^{T}_{\nu}M_{\nu}S_{\nu}R_{1\nu}\simeq\notag\\
&&{\tiny\left(
\begin{array}{ccc}
M_{1}+\frac{1}{\sqrt{2}}M_{3}B_{1}^{T}+\frac{1}{\sqrt{2}}%
B_{1}M_{3}^{T}-B_{1}M_{4}B_{1}^{T} & \frac{1}{\sqrt{2}}M_{3}+\frac{1}{2}%
B_{1}\mathcal{M} & \frac{1}{\sqrt{2}}M_{3}-B_{1}M_{4} \\
\frac{1}{\sqrt{2}}M_{3}^{T}+\frac{1}{2}\mathcal{M} B_{1}^{T} & M_{4} & \frac{1}{2}%
\mathcal{M} -\frac{1}{\sqrt{2}}M_{3}^{T}B_{1} \\
\frac{1}{\sqrt{2}}M_{3}^{T}-M_{4}B_{1}^{T} & \frac{1}{2}\mathcal{M} -\frac{1}{\sqrt{2%
}}B_{1}^{T}M_{3} & -M_{4}-\frac{1}{\sqrt{2}}M_{3}^{T}B_{1}-\frac{1}{\sqrt{2}}%
B_{1}^{T}M_{3}+B_{1}^{T}M_{1}B_{1}%
\end{array}%
\right)}
\label{MT1}
\end{eqnarray}
where the rotation matrix $R_{1\nu}$ is given by:
\begin{eqnarray}
R_{1\nu}=\left(
\begin{array}{ccc}
1-\frac{1}{2}B_{1}B_{1}^{T} & 0 & -B_{1} \\
0 & 1 & 0 \\
B_{1}^{T} & 0 & 1-\frac{1}{2}B_{1}B_{1}^{T}%
\end{array}%
\right)
\end{eqnarray}
The partial diagonalization condition:
\begin{eqnarray}
\left(R^{T}_{1\nu}S^{T}_{\nu}M_{\nu}S_{\nu}R_{1\nu}\right)_{nm}=\left(R^{T}_{1\nu}S^{T}_{\nu}M_{\nu}S_{\nu}R_{1\nu}\right)_{mn}=0,
\hspace{1cm}n=1,2,3\hspace{1cm}m=7,8,9.
\end{eqnarray}
yields the following relation:
\begin{equation}
B_{1}\simeq \frac{1}{\sqrt{2}}M_{3}\left( M_{4}\right) ^{-1}
\end{equation}
Thus, Eq. (\ref{MT1}) takes the form:
\begin{eqnarray}
&&R^{T}_{1\nu}S^{T}_{\nu}M_{\nu}S_{\nu}R_{1\nu}\simeq\left(
\begin{array}{ccc}
M_{1}+\frac{1}{2}M_{3}\left( M_{4}\right) ^{-1}M_{3}^{T} & \frac{1}{\sqrt{2}}%
M_{3} & 0 \\
\frac{1}{\sqrt{2}}M_{3}^{T} & M_{4} & \frac{1}{2}\mathcal{M} -\frac{1}{2}%
M_{3}^{T}M_{3}\left( M_{4}\right) ^{-1} \\
0 & \frac{1}{2}\mathcal{M} -\frac{1}{2}\left( M_{4}\right) ^{-1}M_{3}^{T}M_{3} &
-M_{4}\end{array}
\right)
\label{MT1b}\notag\\
\end{eqnarray}
Now, a third orthogonal transformation is applied under the matrix $M_{\nu}$:
\begin{eqnarray}
&&R^{T}_{2\nu}R^{T}_{1\nu}S^{T}_{\nu}M_{\nu}S_{\nu}R_{1\nu}R_{2\nu}\simeq\\
&&{\tiny\left(
\begin{array}{ccc}
f_{c}\left( M_{1}+\frac{1}{2}M_{3}\left( M_{4}\right) ^{-1}M_{3}^{T}\right)
f_{c}+\frac{1}{\sqrt{2}}f_{c}M_{3}B_{2}^{T}+\frac{1}{\sqrt{2}}%
B_{2}M_{3}^{T}f_{c}+B_{2}M_{4}B_{2}^{T} & \frac{1}{\sqrt{2}}M_{3}+B_{2}M_{4}
& \frac{1}{2}B_{2}\mathcal{M} -\frac{1}{2}B_{2}M_{3}^{T}M_{3}\left( M_{4}\right)
^{-1} \\
\frac{1}{\sqrt{2}}M_{3}^{T}+M_{4}B_{2}^{T} & M_{4}-\frac{1}{\sqrt{2}}\left(
M_{3}^{T}B_{2}+B_{2}^{T}M_{3}\right) & \frac{1}{2}\mathcal{M} -\frac{1}{2}%
M_{3}^{T}M_{3}\left( M_{4}\right) ^{-1} \\
\frac{1}{2}\mathcal{M} B_{2}^{T}-\frac{1}{2}\left( M_{4}\right)
^{-1}M_{3}^{T}M_{3}B_{2}^{T} & \frac{1}{2}\mathcal{M} -\frac{1}{2}\left(
M_{4}\right) ^{-1}M_{3}^{T}M_{3} & -M_{4}%
\end{array}%
\right)}\notag
\label{MT2}
\end{eqnarray}
where
\begin{equation}
f_{s}=B_{2},\hspace{2cm}f_{c}=1-\frac{1}{2}B_{2}B_{2}^{T},
\end{equation}
and the rotation matrix $R_{2\nu}$ is given by:
\begin{eqnarray}
R_{2\nu}=\left(
\begin{array}{ccc}
1-\frac{1}{2}B_{2}B_{2}^{T} & 0 & -B_{2} \\
0 & 1 & 0 \\
B_{2}^{T} & 0 & 1-\frac{1}{2}B_{2}B_{2}^{T}%
\end{array}%
\right)
\end{eqnarray}
The resulting partial diagonalization condition:
\begin{eqnarray}
\left(R^{T}_{2\nu}R^{T}_{1\nu}S^{T}_{\nu}M_{\nu}S_{\nu}R_{1\nu}R_{2\nu}\right)_{nm}=\left(R^{T}_{2\nu}R^{T}_{1\nu}S^{T}_{\nu}M_{\nu}S_{\nu}R_{1\nu}R_{2\nu}\right)_{mn}=0,\hspace{0.5cm}n=1,2,3\hspace{0.5cm}m=4,5,6\notag\\
\end{eqnarray}
yields the following relation:
\begin{equation}
B_{2}\simeq -\frac{1}{\sqrt{2}}M_{3}\left( M_{4}\right) ^{-1}
\end{equation}
Consequently, Eq. (\ref{MT2}) takes the form:
\begin{eqnarray}
&&R^{T}_{2\nu}R^{T}_{1\nu}S^{T}_{\nu}M_{\nu}S_{\nu}R_{1\nu}R_{2\nu}\simeq\notag
\\[3mm]
&&{\tiny\left(
\begin{array}{ccc}
\widetilde{M}_{1} & 0 & -\frac{1}{\sqrt{2}}M_{3}\left( M_{4}\right)
^{-1}\left( \frac{1}{2}\mathcal{M} -\frac{1}{2}M_{3}^{T}M_{3}\left( M_{4}\right)
^{-1}\right) \\
0 & \widetilde{M}_{4} & \frac{1}{2}\mathcal{M} -\frac{1}{2}M_{3}^{T}M_{3}\left(
M_{4}\right) ^{-1} \\
-\frac{1}{\sqrt{2}}\left( \frac{1}{2}\mathcal{M} -\frac{1}{2}\left( M_{4}\right)
^{-1}M_{3}^{T}M_{3}\right) \left( M_{4}\right) ^{-1}M_{3}^{T} & \frac{1}{2}%
\mathcal{M} -\frac{1}{2}\left( M_{4}\right) ^{-1}M_{3}^{T}M_{3} & -M_{4}%
\end{array}%
\right)
}
\notag \\[3mm]
&=&\left(
\begin{array}{ccc}
W & 0 & X \\
0 & Z & Y \\
X^{T} & Y^{T} & -M_{4}%
\end{array}%
\right)
\label{MT2b}
\end{eqnarray}
where
\begin{eqnarray}
W&=&\widetilde{M}_{1}=M_{1}+\frac{1}{16}M_{3}\left( M_{4}\right) ^{-2}M_{3}^{T}M_{3}\left(
M_{4}\right) ^{-1}M_{3}^{T}M_{3}\left( M_{4}\right) ^{-2}M_{3}^{T}\notag\\
Z&=&\widetilde{M}_{4}=M_{4}+\frac{1}{\sqrt{2}}\left( M_{3}^{T}M_{3}\left( M_{4}\right)
^{-1}+\left( M_{4}\right) ^{-1}M_{3}^{T}M_{3}\right)\notag\\
X&=&-\frac{1}{\sqrt{2}}M_{3}\left( M_{4}\right)
^{-1}\left( \frac{1}{2}\mathcal{M} -\frac{1}{2}M_{3}^{T}M_{3}\left( M_{4}\right)
^{-1}\right)\notag\\
Y&=&\frac{1}{2}\mathcal{M} -\frac{1}{2}\left( M_{4}\right) ^{-1}M_{3}^{T}M_{3}
\label{WZXY}
\end{eqnarray}
Then we apply a fourth orthogonal transformation under the matrix $M_{\nu}$, as follows:
\begin{eqnarray}
&&R^{T}_{3\nu}R^{T}_{2\nu}R^{T}_{1\nu}S^{T}_{\nu}M_{\nu}S_{\nu}R_{1\nu}R_{2\nu}R_{3\nu}\notag\\
&&\simeq\left(\begin{array}{ccc}
W+X\left( M_{4}\right) ^{-1}X^{T} & 0 & 0 \\
0 & Z & Y \\
0 & Y^{T} & -M_{4}-X^{T}B_{3}-B_{3}^{T}X+B_{3}^{T}WB_{3}%
\end{array}%
\right)\notag\\
\label{MT3}
\end{eqnarray}
where the rotation matrix $R_{3\nu}$ is given by:
\begin{eqnarray}
R_{3\nu}=\left(
\begin{array}{ccc}
1-\frac{1}{2}B_{3}B_{3}^{T} & 0 & -B_{3} \\
0 & 1 & 0 \\
B_{3}^{T} & 0 & 1-\frac{1}{2}B_{3}B_{3}^{T}%
\end{array}%
\right),\hspace{1cm}B_{3}\simeq X\left( M_{4}\right) ^{-1}
\end{eqnarray}
Consequently, the light active neutrino mass matrix takes the form:
\begin{eqnarray}
M_{1\nu }&=&W+X\left( M_{4}\right) ^{-1}X^{T}\notag\\
&=&M_{1}+\frac{1}{16}M_{3}\left( M_{4}\right)
^{-2}M_{3}^{T}M_{3}\left( M_{4}\right) ^{-1}M_{3}^{T}M_{3}\left(
M_{4}\right) ^{-2}M_{3}^{T}\\
&&+\frac{1}{8}M_{3}\left( M_{4}\right) ^{-1}\left( \mathcal{M} -M_{3}^{T}M_{3}\left(
M_{4}\right) ^{-1}\right) \left( M_{4}\right) ^{-1}\left( \mathcal{M} -\left(
M_{4}\right) ^{-1}M_{3}^{T}M_{3}\right) \left( M_{4}\right) ^{-1}M_{3}^{T}\notag
\end{eqnarray}%
On the other hand, from Eq. (\ref{WZXY}) and considering the hierarchy given by Eq. (\ref{Hierarchymatrices}), it follows that the matrix of Eq.  (\ref{MT3}) is nearly block diagonal, which implies that the sterile neutrino mass matrices are given by:
\begin{eqnarray}
M_{2\nu } &=&-M_{4}  \notag \\
M_{3\nu } &=&M_{4}+\frac{1}{\sqrt{2}}\left( M_{3}^{T}M_{3}\left(
M_{4}\right) ^{-1}+\left( M_{4}\right) ^{-1}M_{3}^{T}M_{3}\right)=M_{4}+\Delta .
\label{M3nu}
\end{eqnarray}

\section{Sterile neutrino mass spectrum}
\label{Sterilespectrum}
In this appendix we compute the sterile neutrino mass spectrum. Our starting point is the fact that the sterile neutrino mass matrices $M_{2\nu }=-M_{4}$ and $M_{3\nu }=M_{4}+\Delta$ satisfy the relation:%
\begin{equation}
M_{4}M_{4}^{T}=\left(
\begin{array}{ccc}
\varepsilon _{1}^{2}+\varepsilon _{2}^{2}+\varepsilon _{3}^{2} & \varepsilon
_{1}d_{1}+\varepsilon _{2}d_{2}+\varepsilon _{3}d_{3} & \varepsilon
_{1}d_{4}+\varepsilon _{2}d_{5}+\varepsilon _{3}d_{6} \\
\varepsilon _{1}d_{1}+\varepsilon _{2}d_{2}+\varepsilon _{3}d_{3} &
d_{1}^{2}+d_{2}^{2}+d_{3}^{2} & d_{1}d_{4}+d_{2}d_{5}+d_{3}d_{6} \\
\varepsilon _{1}d_{4}+\varepsilon _{2}d_{5}+\varepsilon _{3}d_{6} &
d_{1}d_{4}+d_{2}d_{5}+d_{3}d_{6} & d_{4}^{2}+d_{5}^{2}+d_{6}^{2}%
\end{array}%
\right) \frac{v_{\chi }^{2}}{2}\allowbreak ,
\end{equation}
where the subleading $\mathcal{O}\left(\frac{v^{2}}{v_{\chi }}\right)$ corrections presented in $M_{3\nu }$ have been neglected. Then, from the previous expression, it follows that:
\begin{equation}
\det \left( M_{4}M_{4}^{T}\right) =\left( \varepsilon
_{1}d_{2}d_{6}-\varepsilon _{1}d_{3}d_{5}-\varepsilon
_{2}d_{1}d_{6}+\varepsilon _{2}d_{3}d_{4}+\varepsilon
_{3}d_{1}d_{5}-\varepsilon _{3}d_{2}d_{4}\right) ^{2}\allowbreak \frac{%
v_{\chi }^{6}}{8}\allowbreak \allowbreak ,
\end{equation}
Since $\varepsilon _{i}\ll d_{k}$ ($i=1,2,3$ and $k=1,2,\cdots 6$), the mixing angles between the first
generation sterile neutrinos and the second and third generation ones can be
neglected, being of the order of $\frac{\varepsilon _{i}}{d_{k}}$.
Consequently, the masses for the first, second and third generation sterile
neutrinos are respectively given by:
\begin{eqnarray}
m^{(1,2)}_{1}&=&\frac{4\left( \varepsilon _{1}d_{2}d_{6}-\varepsilon
_{1}d_{3}d_{5}-\varepsilon _{2}d_{1}d_{6}+\varepsilon
_{2}d_{3}d_{4}+\varepsilon _{3}d_{1}d_{5}-\varepsilon _{3}d_{2}d_{4}\right)
\allowbreak }{\left( r^{2}-s\right) }\frac{v_{\chi }}{\sqrt{2}},\notag\\
m^{(1,2)}_{2}&=&\frac{1}{2}\left( r-\sqrt{s}\right) \frac{v_{\chi }}{\sqrt{2}},\hspace{1cm}m^{(1,2)}_{3}=\frac{1}{2}\left( r+\sqrt{s}\right) \frac{v_{\chi }}{\sqrt{2}},
\label{sterileneutrinomasses}
\end{eqnarray}
where the superscripts $1$ and $2$ in Eq. (\ref{sterileneutrinomasses}) correspond to the physical neutrino states arising from $M_{2\nu }$ and $M_{3\nu}$, respectively. Furthermore, $r$ and $s$ are given by:
\begin{eqnarray}
r &=&d_{1}^{2}+d_{2}^{2}+d_{3}^{2}+d_{4}^{2}+d_{5}^{2}+d_{6}^{2},\notag \\
s
&=&d_{1}^{4}+2d_{1}^{2}d_{2}^{2}+2d_{1}^{2}d_{3}^{2}+2d_{1}^{2}d_{4}^{2}-2d_{1}^{2}d_{5}^{2}-2d_{1}^{2}d_{6}^{2}+8d_{1}d_{2}d_{4}d_{5}+8d_{1}d_{3}d_{4}d_{6}+d_{2}^{4}+2d_{2}^{2}d_{3}^{2}
\notag\\
&&-2d_{2}^{2}d_{4}^{2}+2d_{2}^{2}d_{5}^{2}-2d_{2}^{2}d_{6}^{2}+8d_{2}d_{3}d_{5}d_{6}+d_{3}^{4}-2d_{3}^{2}d_{4}^{2}-2d_{3}^{2}d_{5}^{2}+2d_{3}^{2}d_{6}^{2}+d_{4}^{4}+2d_{4}^{2}d_{5}^{2}\notag\\
&&+2d_{4}^{2}d_{6}^{2}+d_{5}^{4}+2d_{5}^{2}d_{6}^{2}+d_{6}^{4}
\end{eqnarray}
Since $v_{\chi }\sim \mathcal{O}(10)\mbox{TeV}$, $d_{k}\sim \mathcal{O}(1)$
and $\varepsilon _{i}\sim \mathcal{O}(10^{-2})$, we find $m^{(1,2)}_{1}\sim \mathcal{%
O}(100)\mbox{GeV}$ and $m^{(1,2)}_{2,3}\sim \mathcal{O}(10)\mbox{TeV}$. This shows that the second and third generation of exotic neutrinos arising from $M_{2\nu}$ and $M_{3\nu }$ have $\mathcal{O}(10)\mbox{TeV}$ scale masses, whereas the first generation ones from $M_{2\nu}$ and $M_{3\nu }$ have masses at the electroweak symmetry breaking scale. The $\mathcal{O}(10)\mbox{TeV}$ scale exotic neutrinos of $M_{2\nu}$ have a small splitting of $\sim \frac{v^{2}}{v_{\chi }}$ with the ones of $M_{3\nu }$, as indicated by Eq. (\ref{M3nu}).


\begin{thebibliography}{9}
\bibitem{Aad:2012tfa} 
  G.~Aad {\it et al.} [ATLAS Collaboration],
  Phys.\ Lett.\ B {\bf 716}, 1 (2012)
  doi:10.1016/j.physletb.2012.08.020
  [arXiv:1207.7214 [hep-ex]].



\bibitem{Chatrchyan:2012xdj} 
  S.~Chatrchyan {\it et al.} [CMS Collaboration],
  Phys.\ Lett.\ B {\bf 716}, 30 (2012)
  doi:10.1016/j.physletb.2012.08.021
  [arXiv:1207.7235 [hep-ex]].



\bibitem{Agashe:2014kda} 
  K.~A.~Olive {\it et al.} [Particle Data Group],
  Chin.\ Phys.\ C {\bf 38}, 090001 (2014).
  doi:10.1088/1674-1137/38/9/090001



\bibitem{Patel:2010hr} 
  K.~M.~Patel,
  Phys.\ Lett.\ B {\bf 695}, 225 (2011)
  doi:10.1016/j.physletb.2010.11.024
  [arXiv:1008.5061 [hep-ph]].



\bibitem{Morisi:2011pm} 
  S.~Morisi, K.~M.~Patel and E.~Peinado,
  Phys.\ Rev.\ D {\bf 84}, 053002 (2011)
  doi:10.1103/PhysRevD.84.053002
  [arXiv:1107.0696 [hep-ph]].



\bibitem{Gupta:2011ct} 
  S.~Gupta, A.~S.~Joshipura and K.~M.~Patel,
  Phys.\ Rev.\ D {\bf 85}, 031903 (2012)
  doi:10.1103/PhysRevD.85.031903
  [arXiv:1112.6113 [hep-ph]].



\bibitem{BhupalDev:2012nm} 
  P.~S.~Bhupal Dev, B.~Dutta, R.~N.~Mohapatra and M.~Severson,
  Phys.\ Rev.\ D {\bf 86}, 035002 (2012)
  doi:10.1103/PhysRevD.86.035002
  [arXiv:1202.4012 [hep-ph]].



\bibitem{King:2013hj} 
  S.~F.~King, S.~Morisi, E.~Peinado and J.~W.~F.~Valle,
  Phys.\ Lett.\ B {\bf 724}, 68 (2013)
  doi:10.1016/j.physletb.2013.05.067
  [arXiv:1301.7065 [hep-ph]].



\bibitem{Aranda:2013gga} 
  A.~Aranda, C.~Bonilla, S.~Morisi, E.~Peinado and J.~W.~F.~Valle,
  Phys.\ Rev.\ D {\bf 89}, no. 3, 033001 (2014)
  doi:10.1103/PhysRevD.89.033001
  [arXiv:1307.3553 [hep-ph]].



\bibitem{Gomez-Izquierdo:2013uaa} 
  J.~C.~Gómez-Izquierdo, F.~González-Canales and M.~Mondragon,
  Eur.\ Phys.\ J.\ C {\bf 75}, no. 5, 221 (2015)
  doi:10.1140/epjc/s10052-015-3440-7
  [arXiv:1312.7385 [hep-ph]].



\bibitem{Hernandez:2013dta} 
  A.~E.~Carcamo Hernandez, I.~de Medeiros Varzielas, S.~G.~Kovalenko, H.~Päs and I.~Schmidt,
  Phys.\ Rev.\ D {\bf 88}, no. 7, 076014 (2013)
  doi:10.1103/PhysRevD.88.076014
  [arXiv:1307.6499 [hep-ph]].



\bibitem{Campos:2014lla} 
  M.~D.~Campos, A.~E.~Cárcamo Hernández, S.~Kovalenko, I.~Schmidt and E.~Schumacher,
  Phys.\ Rev.\ D {\bf 90}, no. 1, 016006 (2014)
  doi:10.1103/PhysRevD.90.016006
  [arXiv:1403.2525 [hep-ph]].



\bibitem{Bonilla:2014xla} 
  C.~Bonilla, S.~Morisi, E.~Peinado and J.~W.~F.~Valle,
  Phys.\ Lett.\ B {\bf 742}, 99 (2015)
  doi:10.1016/j.physletb.2015.01.017
  [arXiv:1411.4883 [hep-ph]].



\bibitem{Campos:2014zaa} 
  M.~D.~Campos, A.~E.~Cárcamo Hernández, H.~Päs and E.~Schumacher,
  Phys.\ Rev.\ D {\bf 91}, no. 11, 116011 (2015)
  doi:10.1103/PhysRevD.91.116011
  [arXiv:1408.1652 [hep-ph]].



\bibitem{Arbelaez:2015toa} 
  C.~Arbeláez, A.~E.~Cárcamo Hernández, S.~Kovalenko and I.~Schmidt,
  Phys.\ Rev.\ D {\bf 92}, no. 11, 115015 (2015)
  doi:10.1103/PhysRevD.92.115015
  [arXiv:1507.03852 [hep-ph]].



\bibitem{Hernandez:2015dga} 
  A.~E.~Cárcamo Hernández, I.~de Medeiros Varzielas and E.~Schumacher,
  Phys.\ Rev.\ D {\bf 93}, no. 1, 016003 (2016)
  doi:10.1103/PhysRevD.93.016003
  [arXiv:1509.02083 [hep-ph]].



\bibitem{Hernandez:2015zeh} 
  A.~E.~Cárcamo Hernández, I.~de Medeiros Varzielas and N.~A.~Neill,
  Phys.\ Rev.\ D {\bf 94}, no. 3, 033011 (2016)
  doi:10.1103/PhysRevD.94.033011
  [arXiv:1511.07420 [hep-ph]].



\bibitem{Hernandez:2015hrt} 
  A.~E.~Cárcamo Hernández,
  Eur.\ Phys.\ J.\ C {\bf 76}, no. 9, 503 (2016)
  doi:10.1140/epjc/s10052-016-4351-y
  [arXiv:1512.09092 [hep-ph]].



\bibitem{Joshipura:2015dsa} 
  A.~S.~Joshipura and K.~M.~Patel,
  Phys.\ Lett.\ B {\bf 749}, 159 (2015)
  doi:10.1016/j.physletb.2015.07.062
  [arXiv:1507.01235 [hep-ph]].



\bibitem{Bjorkeroth:2015uou} 
  F.~Björkeroth, F.~J.~de Anda, I.~de Medeiros Varzielas and S.~F.~King,
  Phys.\ Rev.\ D {\bf 94}, no. 1, 016006 (2016)
  doi:10.1103/PhysRevD.94.016006
  [arXiv:1512.00850 [hep-ph]].



\bibitem{Bjorkeroth:2015ora} 
  F.~Björkeroth, F.~J.~de Anda, I.~de Medeiros Varzielas and S.~F.~King,
  JHEP {\bf 1506}, 141 (2015)
  doi:10.1007/JHEP06(2015)141
  [arXiv:1503.03306 [hep-ph]].



\bibitem{Varzielas:2015aua} 
  I.~de Medeiros Varzielas,
  JHEP {\bf 1508}, 157 (2015)
  doi:10.1007/JHEP08(2015)157
  [arXiv:1507.00338 [hep-ph]].



\bibitem{Chen:2015jta} 
  P.~Chen, G.~J.~Ding, A.~D.~Rojas, C.~A.~Vaquera-Araujo and J.~W.~F.~Valle,
  JHEP {\bf 1601}, 007 (2016)
  doi:10.1007/JHEP01(2016)007
  [arXiv:1509.06683 [hep-ph]].



\bibitem{Ma:2015fpa} 
  E.~Ma,
  Phys.\ Lett.\ B {\bf 752}, 198 (2016)
  doi:10.1016/j.physletb.2015.11.049
  [arXiv:1510.02501 [hep-ph]].



\bibitem{Ma:2016nkf} 
  E.~Ma,
  Phys.\ Lett.\ B {\bf 755}, 348 (2016)
  doi:10.1016/j.physletb.2016.02.032
  [arXiv:1601.00138 [hep-ph]].



\bibitem{Arbelaez:2016mhg} 
  C.~Arbeláez, A.~E.~Cárcamo Hernández, S.~Kovalenko and I.~Schmidt,
  Eur.\ Phys.\ J.\ C {\bf 77}, no. 6, 422 (2017)
  doi:10.1140/epjc/s10052-017-4948-9
  [arXiv:1602.03607 [hep-ph]].



\bibitem{Pasquini:2016kwk} 
  P.~Pasquini, S.~C.~Chuliá and J.~W.~F.~Valle,
  Phys.\ Rev.\ D {\bf 95}, no. 9, 095030 (2017)
  doi:10.1103/PhysRevD.95.095030
  [arXiv:1610.05962 [hep-ph]].



\bibitem{Carballo-Perez:2016ooy} 
  B.~Carballo-Perez, E.~Peinado and S.~Ramos-Sanchez,
  JHEP {\bf 1612}, 131 (2016)
  doi:10.1007/JHEP12(2016)131
  [arXiv:1607.06812 [hep-ph]].



\bibitem{Vien:2016qbb} 
  V.~V.~Vien and H.~N.~Long,
  arXiv:1609.03895 [hep-ph].



\bibitem{Chulia:2016giq} 
  S.~Centelles Chuliá, R.~Srivastava and J.~W.~F.~Valle,
  Phys.\ Lett.\ B {\bf 761}, 431 (2016)
  doi:10.1016/j.physletb.2016.08.028
  [arXiv:1606.06904 [hep-ph]].



\bibitem{Penedo:2017vtf} 
  J.~T.~Penedo, S.~T.~Petcov and A.~V.~Titov,
  JHEP {\bf 1712}, 022 (2017)
  doi:10.1007/JHEP12(2017)022
  [arXiv:1705.00309 [hep-ph]].



\bibitem{Cruz:2017add} 
  A.~A.~Cruz and M.~Mondragón,
  arXiv:1701.07929 [hep-ph].



\bibitem{Vien:2013zra} 
  V.~V.~Vien and H.~N.~Long,
  Int.\ J.\ Mod.\ Phys.\ A {\bf 28}, 1350159 (2013)
  doi:10.1142/S0217751X13501595
  [arXiv:1312.5034 [hep-ph]].



\bibitem{Hernandez:2013hea} 
  A.~E.~Cárcamo Hernández, R.~Martinez and F.~Ochoa,
  Eur.\ Phys.\ J.\ C {\bf 76}, no. 11, 634 (2016)
  doi:10.1140/epjc/s10052-016-4480-3
  [arXiv:1309.6567 [hep-ph]].



\bibitem{Vien:2014pta} 
  V.~V.~Vien and H.~N.~Long,
  Int.\ J.\ Mod.\ Phys.\ A {\bf 30}, no. 21, 1550117 (2015)
  doi:10.1142/S0217751X15501171
  [arXiv:1405.4665 [hep-ph]].



\bibitem{Vien:2014gza} 
  V.~V.~Vien and H.~N.~Long,
  JHEP {\bf 1404}, 133 (2014)
  doi:10.1007/JHEP04(2014)133
  [arXiv:1402.1256 [hep-ph]].



\bibitem{Vien:2014vka} 
  V.~V.~Vien and H.~N.~Long,
  Zh.\ Eksp.\ Teor.\ Fiz.\  {\bf 145}, 991 (2014)
  [J.\ Exp.\ Theor.\ Phys.\  {\bf 118}, no. 6, 869 (2014)]
  doi:10.7868/S0044451014060044, 10.1134/S1063776114050173
  [arXiv:1404.6119 [hep-ph]].



\bibitem{Vien:2014ica} 
  V.~V.~Vien and H.~N.~Long,
  J.\ Korean Phys.\ Soc.\  {\bf 66}, no. 12, 1809 (2015)
  doi:10.3938/jkps.66.1809
  [arXiv:1408.4333 [hep-ph]].



\bibitem{Hernandez:2014lpa} 
  A.~E.~Cárcamo Hernández, E.~Cataño Mur and R.~Martinez,
  Phys.\ Rev.\ D {\bf 90}, no. 7, 073001 (2014)
  doi:10.1103/PhysRevD.90.073001
  [arXiv:1407.5217 [hep-ph]].



\bibitem{Hernandez:2014vta} 
  A.~E.~Cárcamo Hernández, R.~Martinez and J.~Nisperuza,
  Eur.\ Phys.\ J.\ C {\bf 75}, no. 2, 72 (2015)
  doi:10.1140/epjc/s10052-015-3278-z
  [arXiv:1401.0937 [hep-ph]].



\bibitem{VanVien:2015xha} 
  V.~V.~Vien, H.~N.~Long and D.~P.~Khoi,
  Int.\ J.\ Mod.\ Phys.\ A {\bf 30}, no. 17, 1550102 (2015)
  doi:10.1142/S0217751X1550102X
  [arXiv:1506.06063 [hep-ph]].



\bibitem{Hernandez:2015tna} 
  A.~E.~Cárcamo Hernández and R.~Martinez,
  Nucl.\ Phys.\ B {\bf 905}, 337 (2016)
  doi:10.1016/j.nuclphysb.2016.02.025
  [arXiv:1501.05937 [hep-ph]].



\bibitem{Hernandez:2015cra} 
  A.~E.~Cárcamo Hernández and R.~Martinez,
  J.\ Phys.\ G {\bf 43}, no. 4, 045003 (2016)
  doi:10.1088/0954-3899/43/4/045003
  [arXiv:1501.07261 [hep-ph]].



\bibitem{Vien:2016tmh} 
  V.~V.~Vien, A.~E.~Cárcamo Hernández and H.~N.~Long,
  Nucl.\ Phys.\ B {\bf 913}, 792 (2016)
  doi:10.1016/j.nuclphysb.2016.10.010
  [arXiv:1601.03300 [hep-ph]].



\bibitem{Hernandez:2016eod} 
  A.~E.~Cárcamo Hernández, H.~N.~Long and V.~V.~Vien,
  Eur.\ Phys.\ J.\ C {\bf 76}, no. 5, 242 (2016)
  doi:10.1140/epjc/s10052-016-4074-0
  [arXiv:1601.05062 [hep-ph]].



\bibitem{Camargo-Molina:2016yqm} 
  J.~E.~Camargo-Molina, A.~P.~Morais, A.~Ordell, R.~Pasechnik, M.~O.~P.~Sampaio and J.~Wessén,
  Phys.\ Rev.\ D {\bf 95}, no. 7, 075031 (2017)
  doi:10.1103/PhysRevD.95.075031
  [arXiv:1610.03642 [hep-ph]].



\bibitem{Camargo-Molina:2016bwm} 
  J.~E.~Camargo-Molina, A.~P.~Morais, R.~Pasechnik and J.~Wessén,
  JHEP {\bf 1609}, 129 (2016)
  doi:10.1007/JHEP09(2016)129
  [arXiv:1606.03492 [hep-ph]].



\bibitem{Gomez-Izquierdo:2017rxi} 
  J.~C.~Gómez-Izquierdo,
  Eur.\ Phys.\ J.\ C {\bf 77}, no. 8, 551 (2017)
  doi:10.1140/epjc/s10052-017-5094-0
  [arXiv:1701.01747 [hep-ph]].



\bibitem{Gomez-Izquierdo:2017med} 
  J.~C.~Gómez-Izquierdo, F.~Gonzalez-Canales and M.~Mondragón,
  Int.\ J.\ Mod.\ Phys.\ A {\bf 32}, no. 28-29, 1750171 (2017)
  doi:10.1142/S0217751X17501718
  [arXiv:1705.06324 [hep-ph]].



\bibitem{Chattopadhyay:2017zvs} 
  P.~Chattopadhyay and K.~M.~Patel,
  Nucl.\ Phys.\ B {\bf 921}, 487 (2017)
  doi:10.1016/j.nuclphysb.2017.06.008
  [arXiv:1703.09541 [hep-ph]].



\bibitem{CarcamoHernandez:2017kra} 
  A.~E.~Cárcamo Hernández and H.~N.~Long,
  J.\ Phys.\ G {\bf 45}, no. 4, 045001 (2018)
  doi:10.1088/1361-6471/aaace7
  [arXiv:1705.05246 [hep-ph]].



\bibitem{CarcamoHernandez:2017owh} 
  A.~E.~Cárcamo Hernández, S.~Kovalenko, J.~W.~F.~Valle and C.~A.~Vaquera-Araujo,
  JHEP {\bf 1707}, 118 (2017)
  doi:10.1007/JHEP07(2017)118
  [arXiv:1705.06320 [hep-ph]].



\bibitem{Bernal:2017xat} 
  N.~Bernal, A.~E.~Cárcamo Hernández, I.~de Medeiros Varzielas and S.~Kovalenko,
  JHEP {\bf 1805}, 053 (2018)
  doi:10.1007/JHEP05(2018)053
  [arXiv:1712.02792 [hep-ph]].



\bibitem{CarcamoHernandez:2018iel} 
  A.~E.~Cárcamo Hernández, H.~N.~Long and V.~V.~Vien,
  arXiv:1803.01636 [hep-ph].



\bibitem{deAnda:2018oik} 
  F.~J.~de Anda and S.~F.~King,
  arXiv:1803.04978 [hep-ph].



\bibitem{CarcamoHernandez:2018aon} 
  A.~E.~Cárcamo Hernández and S.~F.~King,
  arXiv:1803.07367 [hep-ph].



\bibitem{Altarelli:2002hx} 
  G.~Altarelli and F.~Feruglio,
  Springer Tracts Mod.\ Phys.\  {\bf 190}, 169 (2003)
  [hep-ph/0206077].



\bibitem{Altarelli:2010gt} 
  G.~Altarelli and F.~Feruglio,
  Rev.\ Mod.\ Phys.\  {\bf 82}, 2701 (2010)
  doi:10.1103/RevModPhys.82.2701
  [arXiv:1002.0211 [hep-ph]].



\bibitem{Ishimori:2010au} 
  H.~Ishimori, T.~Kobayashi, H.~Ohki, Y.~Shimizu, H.~Okada and M.~Tanimoto,
  Prog.\ Theor.\ Phys.\ Suppl.\  {\bf 183}, 1 (2010)
  doi:10.1143/PTPS.183.1
  [arXiv:1003.3552 [hep-th]].



\bibitem{King:2013eh} 
  S.~F.~King and C.~Luhn,
  Rept.\ Prog.\ Phys.\  {\bf 76}, 056201 (2013)
  doi:10.1088/0034-4885/76/5/056201
  [arXiv:1301.1340 [hep-ph]].



\bibitem{King:2014nza} 
  S.~F.~King, A.~Merle, S.~Morisi, Y.~Shimizu and M.~Tanimoto,
  New J.\ Phys.\  {\bf 16}, 045018 (2014)
  doi:10.1088/1367-2630/16/4/045018
  [arXiv:1402.4271 [hep-ph]].



\bibitem{King:2017guk} 
  S.~F.~King,
  Prog.\ Part.\ Nucl.\ Phys.\  {\bf 94}, 217 (2017)
  doi:10.1016/j.ppnp.2017.01.003
  [arXiv:1701.04413 [hep-ph]].



\bibitem{Fritzsch:1977za} 
  H.~Fritzsch,
  Phys.\ Lett.\  {\bf 70B}, 436 (1977).
  doi:10.1016/0370-2693(77)90408-7



\bibitem{Fukuyama:1997ky} 
  T.~Fukuyama and H.~Nishiura,
  hep-ph/9702253.



\bibitem{Du:1992iy} 
  D.~s.~Du and Z.~z.~Xing,
  Phys.\ Rev.\ D {\bf 48}, 2349 (1993).
  doi:10.1103/PhysRevD.48.2349



\bibitem{Barbieri:1994kw} 
  R.~Barbieri, G.~R.~Dvali, A.~Strumia, Z.~Berezhiani and L.~J.~Hall,
  Nucl.\ Phys.\ B {\bf 432}, 49 (1994)
  doi:10.1016/0550-3213(94)90593-2
  [hep-ph/9405428].



\bibitem{Peccei:1995fg} 
  R.~D.~Peccei and K.~Wang,
  Phys.\ Rev.\ D {\bf 53}, 2712 (1996)
  doi:10.1103/PhysRevD.53.2712
  [hep-ph/9509242].



\bibitem{Fritzsch:1999ee} 
  H.~Fritzsch and Z.~z.~Xing,
  Prog.\ Part.\ Nucl.\ Phys.\  {\bf 45}, 1 (2000)
  doi:10.1016/S0146-6410(00)00102-2
  [hep-ph/9912358].



\bibitem{Roberts:2001zy} 
  R.~G.~Roberts, A.~Romanino, G.~G.~Ross and L.~Velasco-Sevilla,
  Nucl.\ Phys.\ B {\bf 615}, 358 (2001)
  doi:10.1016/S0550-3213(01)00408-4
  [hep-ph/0104088].



\bibitem{Nishiura:2002ei} 
  H.~Nishiura, K.~Matsuda, T.~Kikuchi and T.~Fukuyama,
  Phys.\ Rev.\ D {\bf 65}, 097301 (2002)
  doi:10.1103/PhysRevD.65.097301
  [hep-ph/0202189].



\bibitem{deMedeirosVarzielas:2005ax} 
  I.~de Medeiros Varzielas and G.~G.~Ross,
  Nucl.\ Phys.\ B {\bf 733}, 31 (2006)
  doi:10.1016/j.nuclphysb.2005.10.039
  [hep-ph/0507176].



\bibitem{Carcamo:2006dp} 
  A.~E.~Carcamo Hernandez, R.~Martinez and J.~A.~Rodriguez,
  Eur.\ Phys.\ J.\ C {\bf 50}, 935 (2007)
  doi:10.1140/epjc/s10052-007-0264-0
  [hep-ph/0606190].

\bibitem{Kajiyama:2007gx} 
  Y.~Kajiyama, M.~Raidal and A.~Strumia,
  Phys.\ Rev.\ D {\bf 76}, 117301 (2007)
  doi:10.1103/PhysRevD.76.117301
  [arXiv:0705.4559 [hep-ph]].



\bibitem{CarcamoHernandez:2010im} 
  A.~E.~Carcamo Hernandez and R.~Rahman,
  Rev.\ Mex.\ Fis.\  {\bf 62}, no. 2, 100 (2016)
  [arXiv:1007.0447 [hep-ph]].



\bibitem{Branco:2010tx} 
  G.~C.~Branco, D.~Emmanuel-Costa and C.~Simoes,
  Phys.\ Lett.\ B {\bf 690}, 62 (2010)
  doi:10.1016/j.physletb.2010.05.009
  [arXiv:1001.5065 [hep-ph]].



\bibitem{Leser:2011fz} 
  P.~Leser and H.~Pas,
  Phys.\ Rev.\ D {\bf 84}, 017303 (2011)
  doi:10.1103/PhysRevD.84.017303
  [arXiv:1104.2448 [hep-ph]].



\bibitem{Gupta:2012dma} 
  M.~Gupta and G.~Ahuja,
  Int.\ J.\ Mod.\ Phys.\ A {\bf 27}, 1230033 (2012)
  doi:10.1142/S0217751X12300335
  [arXiv:1302.4823 [hep-ph]].



\bibitem{Dev:2012sg} 
  P.~S.~B.~Dev and A.~Pilaftsis,
  Phys.\ Rev.\ D {\bf 86}, 113001 (2012)
  doi:10.1103/PhysRevD.86.113001
  [arXiv:1209.4051 [hep-ph]].



\bibitem{CarcamoHernandez:2012xy} 
  A.~E.~Carcamo Hernandez, C.~O.~Dib, N.~Neill H and A.~R.~Zerwekh,
  JHEP {\bf 1202}, 132 (2012)
  doi:10.1007/JHEP02(2012)132
  [arXiv:1201.0878 [hep-ph]].



\bibitem{Hernandez:2013mcf} 
  A.~E.~Carcamo Hernandez, R.~Martinez and F.~Ochoa,
  Phys.\ Rev.\ D {\bf 87}, no. 7, 075009 (2013)
  doi:10.1103/PhysRevD.87.075009
  [arXiv:1302.1757 [hep-ph]].



\bibitem{Pas:2014bra} 
  H.~Päs and E.~Schumacher,
  Phys.\ Rev.\ D {\bf 89}, no. 9, 096010 (2014)
  doi:10.1103/PhysRevD.89.096010
  [arXiv:1401.2328 [hep-ph]].



\bibitem{Hernandez:2014hka} 
  A.~E.~Carcamo Hernandez, S.~Kovalenko and I.~Schmidt,
  arXiv:1411.2913 [hep-ph].



\bibitem{Hernandez:2014zsa} 
  A.~E.~Cárcamo Hernández and I.~de Medeiros Varzielas,
  J.\ Phys.\ G {\bf 42}, no. 6, 065002 (2015)
  doi:10.1088/0954-3899/42/6/065002
  [arXiv:1410.2481 [hep-ph]].



\bibitem{Nishiura:2014psa} 
  H.~Nishiura and T.~Fukuyama,
  Mod.\ Phys.\ Lett.\ A {\bf 29}, 0147 (2014)
  doi:10.1142/S0217732314501478
  [arXiv:1405.2416 [hep-ph]].



\bibitem{Frank:2014aca} 
  M.~Frank, C.~Hamzaoui, N.~Pourtolami and M.~Toharia,
  Phys.\ Lett.\ B {\bf 742}, 178 (2015)
  doi:10.1016/j.physletb.2015.01.025
  [arXiv:1406.2331 [hep-ph]].



\bibitem{Ghosal:2015lwa} 
  A.~Ghosal and R.~Samanta,
  JHEP {\bf 1505}, 077 (2015)
  doi:10.1007/JHEP05(2015)077
  [arXiv:1501.00916 [hep-ph]].



\bibitem{Sinha:2015ooa} 
  R.~Sinha, R.~Samanta and A.~Ghosal,
  Phys.\ Lett.\ B {\bf 759}, 206 (2016)
  doi:10.1016/j.physletb.2016.05.080
  [arXiv:1508.05227 [hep-ph]].



\bibitem{Nishiura:2015qia} 
  H.~Nishiura and T.~Fukuyama,
  Phys.\ Lett.\ B {\bf 753}, 57 (2016)
  doi:10.1016/j.physletb.2015.11.080
  [arXiv:1510.01035 [hep-ph]].



\bibitem{Samanta:2015oqa} 
  R.~Samanta and A.~Ghosal,
  Nucl.\ Phys.\ B {\bf 911}, 846 (2016)
  doi:10.1016/j.nuclphysb.2016.08.036
  [arXiv:1507.02582 [hep-ph]].



\bibitem{Gautam:2015kya} 
  R.~R.~Gautam, M.~Singh and M.~Gupta,
  Phys.\ Rev.\ D {\bf 92}, no. 1, 013006 (2015)
  doi:10.1103/PhysRevD.92.013006
  [arXiv:1506.04868 [hep-ph]].



\bibitem{Pas:2015hca} 
  H.~Päs and E.~Schumacher,
  Phys.\ Rev.\ D {\bf 92}, no. 11, 114025 (2015)
  doi:10.1103/PhysRevD.92.114025
  [arXiv:1510.08757 [hep-ph]].



\bibitem{Gupta:2013yha} 
  M.~Gupta and G.~Ahuja,
  Int.\ J.\ Mod.\ Phys.\ A {\bf 27}, 1230033 (2012)
  doi:10.1142/S0217751X12300335
  [arXiv:1302.4823 [hep-ph]].



\bibitem{Wang:2015saa} 
  W.~Wang and Z.~L.~Han,
  Phys.\ Rev.\ D {\bf 92}, 095001 (2015)
  doi:10.1103/PhysRevD.92.095001
  [arXiv:1508.00706 [hep-ph]].



\bibitem{vonGersdorff:2017iym} 
  G.~von Gersdorff,
  JHEP {\bf 1709}, 094 (2017)
  doi:10.1007/JHEP09(2017)094
  [arXiv:1705.05430 [hep-ph]].



\bibitem{Georgi:1978bv} 
  H.~Georgi and A.~Pais,
  Phys.\ Rev.\ D {\bf 19}, 2746 (1979).
  doi:10.1103/PhysRevD.19.2746

\bibitem{Valle:1983dk} 
  J.~W.~F.~Valle and M.~Singer,
  Phys.\ Rev.\ D {\bf 28}, 540 (1983).
  doi:10.1103/PhysRevD.28.540



\bibitem{Pisano:1991ee} 
  F.~Pisano and V.~Pleitez,
  Phys.\ Rev.\ D {\bf 46}, 410 (1992)
  doi:10.1103/PhysRevD.46.410
  [hep-ph/9206242].



\bibitem{Foot:1992rh} 
  R.~Foot, O.~F.~Hernandez, F.~Pisano and V.~Pleitez,
  Phys.\ Rev.\ D {\bf 47}, 4158 (1993)
  doi:10.1103/PhysRevD.47.4158
  [hep-ph/9207264].



\bibitem{Frampton:1992wt} 
  P.~H.~Frampton,
  Phys.\ Rev.\ Lett.\  {\bf 69}, 2889 (1992).
  doi:10.1103/PhysRevLett.69.2889



\bibitem{Hoang:1996gi} 
  H.~N.~Long,
  Phys.\ Rev.\ D {\bf 54}, 4691 (1996)
  doi:10.1103/PhysRevD.54.4691
  [hep-ph/9607439].



\bibitem{Hoang:1995vq} 
  H.~N.~Long,
  Phys.\ Rev.\ D {\bf 53}, 437 (1996)
  doi:10.1103/PhysRevD.53.437
  [hep-ph/9504274].



\bibitem{Foot:1994ym} 
  R.~Foot, H.~N.~Long and T.~A.~Tran,
  Phys.\ Rev.\ D {\bf 50}, no. 1, R34 (1994)
  doi:10.1103/PhysRevD.50.R34
  [hep-ph/9402243].



\bibitem{Boucenna:2015pav} 
  S.~M.~Boucenna, S.~Morisi and A.~Vicente,
  Phys.\ Rev.\ D {\bf 93}, no. 11, 115008 (2016)
  doi:10.1103/PhysRevD.93.115008
  [arXiv:1512.06878 [hep-ph]].



\bibitem{Hernandez:2015ywg} 
  A.~E.~C.~Hernández and I.~Nišandžić,
  Eur.\ Phys.\ J.\ C {\bf 76}, no. 7, 380 (2016)
  doi:10.1140/epjc/s10052-016-4230-6
  [arXiv:1512.07165 [hep-ph]].



\bibitem{Fonseca:2016tbn} 
  R.~M.~Fonseca and M.~Hirsch,
  JHEP {\bf 1608}, 003 (2016)
  doi:10.1007/JHEP08(2016)003
  [arXiv:1606.01109 [hep-ph]].
  
  
\bibitem{Hati:2017aez} 
  C.~Hati, S.~Patra, M.~Reig, J.~W.~F.~Valle and C.~A.~Vaquera-Araujo,
  Phys.\ Rev.\ D {\bf 96}, no. 1, 015004 (2017)
  doi:10.1103/PhysRevD.96.015004
  [arXiv:1703.09647 [hep-ph]].



\bibitem{deSousaPires:1998jc} 
  C.~A.~de Sousa Pires and O.~P.~Ravinez,
  Phys.\ Rev.\ D {\bf 58}, 035008 (1998)
  [Phys.\ Rev.\ D {\bf 58}, 35008 (1998)]
  doi:10.1103/PhysRevD.58.035008
  [hep-ph/9803409].



\bibitem{VanDong:2005ux} 
  P.~V.~Dong and H.~N.~Long,
  Int.\ J.\ Mod.\ Phys.\ A {\bf 21}, 6677 (2006)
  doi:10.1142/S0217751X06035191
  [hep-ph/0507155].



\bibitem{Montero:1998yw} 
  J.~C.~Montero, V.~Pleitez and O.~Ravinez,
  Phys.\ Rev.\ D {\bf 60}, 076003 (1999)
  doi:10.1103/PhysRevD.60.076003
  [hep-ph/9811280].



\bibitem{Montero:2005yb} 
  J.~C.~Montero, C.~C.~Nishi, V.~Pleitez, O.~Ravinez and M.~C.~Rodriguez,
  Phys.\ Rev.\ D {\bf 73}, 016003 (2006)
  doi:10.1103/PhysRevD.73.016003
  [hep-ph/0511100].



\bibitem{Pal:1994ba} 
  P.~B.~Pal,
  Phys.\ Rev.\ D {\bf 52}, 1659 (1995)
  doi:10.1103/PhysRevD.52.1659
  [hep-ph/9411406].



\bibitem{Dias:2002gg} 
  A.~G.~Dias, V.~Pleitez and M.~D.~Tonasse,
  Phys.\ Rev.\ D {\bf 67}, 095008 (2003)
  doi:10.1103/PhysRevD.67.095008
  [hep-ph/0211107].



\bibitem{Dias:2003zt} 
  A.~G.~Dias and V.~Pleitez,
  Phys.\ Rev.\ D {\bf 69}, 077702 (2004)
  doi:10.1103/PhysRevD.69.077702
  [hep-ph/0308037].



\bibitem{Dias:2003iq} 
  A.~G.~Dias, C.~A.~de S. Pires and P.~S.~Rodrigues da Silva,
  Phys.\ Rev.\ D {\bf 68}, 115009 (2003)
  doi:10.1103/PhysRevD.68.115009
  [hep-ph/0309058].



\bibitem{Mizukoshi:2010ky} 
  J.~K.~Mizukoshi, C.~A.~de S.Pires, F.~S.~Queiroz and P.~S.~Rodrigues da Silva,
  Phys.\ Rev.\ D {\bf 83}, 065024 (2011)
  doi:10.1103/PhysRevD.83.065024
  [arXiv:1010.4097 [hep-ph]].



\bibitem{Dias:2010vt} 
  A.~G.~Dias, C.~A.~de S.Pires and P.~S.~Rodrigues da Silva,
  Phys.\ Rev.\ D {\bf 82}, 035013 (2010)
  doi:10.1103/PhysRevD.82.035013
  [arXiv:1003.3260 [hep-ph]].



\bibitem{Alvares:2012qv} 
  J.~D.~Ruiz-Alvarez, C.~A.~de S.Pires, F.~S.~Queiroz, D.~Restrepo and P.~S.~Rodrigues da Silva,
  Phys.\ Rev.\ D {\bf 86}, 075011 (2012)
  doi:10.1103/PhysRevD.86.075011
  [arXiv:1206.5779 [hep-ph]].



\bibitem{Cogollo:2014jia} 
  D.~Cogollo, A.~X.~Gonzalez-Morales, F.~S.~Queiroz and P.~R.~Teles,
  JCAP {\bf 1411}, no. 11, 002 (2014)
  doi:10.1088/1475-7516/2014/11/002
  [arXiv:1402.3271 [hep-ph]].



\bibitem{daSilva:2014qba} 
  P.~S.~Rodrigues da Silva,
  Phys.\ Int.\  {\bf 7}, no. 1, 15 (2016)
  doi:10.3844/pisp.2016.15.27
  [arXiv:1412.8633 [hep-ph]].



\bibitem{Kitabayashi:2002jd} 
  T.~Kitabayashi and M.~Yasue,
  Phys.\ Rev.\ D {\bf 67}, 015006 (2003)
  doi:10.1103/PhysRevD.67.015006
  [hep-ph/0209294].



\bibitem{Sen:2007vx} 
  S.~Sen,
  Phys.\ Rev.\ D {\bf 76}, 115020 (2007)
  doi:10.1103/PhysRevD.76.115020
  [arXiv:0710.2734 [hep-ph]].



\bibitem{Dong:2010zu} 
  P.~V.~Dong, H.~N.~Long, D.~V.~Soa and V.~V.~Vien,
  Eur.\ Phys.\ J.\ C {\bf 71}, 1544 (2011)
  doi:10.1140/epjc/s10052-011-1544-2
  [arXiv:1009.2328 [hep-ph]].



\bibitem{Dong:2011vb} 
  P.~V.~Dong, H.~N.~Long, C.~H.~Nam and V.~V.~Vien,
  Phys.\ Rev.\ D {\bf 85}, 053001 (2012)
  doi:10.1103/PhysRevD.85.053001
  [arXiv:1111.6360 [hep-ph]].



\bibitem{Kitabayashi:2000wh} 
  T.~Kitabayashi and M.~Yasue,
  hep-ph/0006040.



\bibitem{Kitabayashi:2000nq} 
  T.~Kitabayashi and M.~Yasue,
  Phys.\ Rev.\ D {\bf 63}, 095002 (2001)
  doi:10.1103/PhysRevD.63.095002
  [hep-ph/0010087].



\bibitem{Kitabayashi:2001id} 
  T.~Kitabayashi and M.~Yasue,
  Phys.\ Lett.\ B {\bf 508}, 85 (2001)
  doi:10.1016/S0370-2693(01)00397-5
  [hep-ph/0102228].



\bibitem{Chang:2006aa} 
  D.~Chang and H.~N.~Long,
  Phys.\ Rev.\ D {\bf 73}, 053006 (2006)
  doi:10.1103/PhysRevD.73.053006
  [hep-ph/0603098].



\bibitem{Dong:2006gx} 
  P.~V.~Dong, D.~T.~Huong, T.~T.~Huong and H.~N.~Long,
  Phys.\ Rev.\ D {\bf 74}, 053003 (2006)
  doi:10.1103/PhysRevD.74.053003
  [hep-ph/0607291].



\bibitem{Dong:2008sw} 
  P.~V.~Dong and H.~N.~Long,
  Phys.\ Rev.\ D {\bf 77}, 057302 (2008)
  doi:10.1103/PhysRevD.77.057302
  [arXiv:0801.4196 [hep-ph]].



\bibitem{Martinez:2011iq} 
  R.~Martinez, F.~Ochoa and P.~Fonseca,
  arXiv:1105.4623 [hep-ph].



\bibitem{Huong:2012pg} 
  D.~T.~Huong, L.~T.~Hue, M.~C.~Rodriguez and H.~N.~Long,
  Nucl.\ Phys.\ B {\bf 870}, 293 (2013)
  doi:10.1016/j.nuclphysb.2013.01.016
  [arXiv:1210.6776 [hep-ph]].



\bibitem{Boucenna:2014ela} 
  S.~M.~Boucenna, S.~Morisi and J.~W.~F.~Valle,
  Phys.\ Rev.\ D {\bf 90}, no. 1, 013005 (2014)
  doi:10.1103/PhysRevD.90.013005
  [arXiv:1405.2332 [hep-ph]].



\bibitem{Okada:2015bxa} 
  H.~Okada, N.~Okada and Y.~Orikasa,
  Phys.\ Rev.\ D {\bf 93}, no. 7, 073006 (2016)
  doi:10.1103/PhysRevD.93.073006
  [arXiv:1504.01204 [hep-ph]].



\bibitem{Pires:2014xsa} 
  C.~A.~d.~S.~Pires,
  doi:10.3844/pisp.2015.33.41
  arXiv:1412.1002 [hep-ph].



\bibitem{Catano:2012kw} 
  M.~E.~Catano, R.~Martinez and F.~Ochoa,
  Phys.\ Rev.\ D {\bf 86}, 073015 (2012)
  doi:10.1103/PhysRevD.86.073015
  [arXiv:1206.1966 [hep-ph]].



\bibitem{Reig:2016ewy} 
  M.~Reig, J.~W.~F.~Valle and C.~A.~Vaquera-Araujo,
  Phys.\ Rev.\ D {\bf 94}, no. 3, 033012 (2016)
  doi:10.1103/PhysRevD.94.033012
  [arXiv:1606.08499 [hep-ph]].



\bibitem{CarcamoHernandez:2016pdu} 
  A.~E.~Cárcamo Hernández, S.~Kovalenko and I.~Schmidt,
  JHEP {\bf 1702}, 125 (2017)
  doi:10.1007/JHEP02(2017)125
  [arXiv:1611.09797 [hep-ph]].



\bibitem{Singer:1980sw} 
  M.~Singer, J.~W.~F.~Valle and J.~Schechter,
  Phys.\ Rev.\ D {\bf 22}, 738 (1980).
  doi:10.1103/PhysRevD.22.738



\bibitem{Montero:1992jk} 
  J.~C.~Montero, F.~Pisano and V.~Pleitez,
  Phys.\ Rev.\ D {\bf 47}, 2918 (1993)
  doi:10.1103/PhysRevD.47.2918
  [hep-ph/9212271].



\bibitem{Sanchez:2001ua} 
  L.~A.~Sanchez, W.~A.~Ponce and R.~Martinez,
  Phys.\ Rev.\ D {\bf 64}, 075013 (2001)
  doi:10.1103/PhysRevD.64.075013
  [hep-ph/0103244].



\bibitem{Ponce:2001jn} 
  W.~A.~Ponce, J.~B.~Florez and L.~A.~Sanchez,
  Int.\ J.\ Mod.\ Phys.\ A {\bf 17}, 643 (2002)
  doi:10.1142/S0217751X02005815
  [hep-ph/0103100].



\bibitem{Ponce:2002sg} 
  W.~A.~Ponce, Y.~Giraldo and L.~A.~Sanchez,
  Phys.\ Rev.\ D {\bf 67}, 075001 (2003)
  doi:10.1103/PhysRevD.67.075001
  [hep-ph/0210026].



\bibitem{Martinez:2006gb} 
  R.~Martinez and F.~Ochoa,
  Eur.\ Phys.\ J.\ C {\bf 51}, 701 (2007)
  doi:10.1140/epjc/s10052-007-0307-6
  [hep-ph/0606173].



\bibitem{Hue:2015mna}  
L.~T.~Hue and L.~D.~Ninh,
  Mod.\ Phys.\ Lett.\ A {\bf 31}, no. 10, 1650062 (2016)
  doi:10.1142/S0217732316500620
  [arXiv:1510.00302 [hep-ph]].



\bibitem{Diaz:2004fs} 
  R.~A.~Diaz, R.~Martinez and F.~Ochoa,
  Phys.\ Rev.\ D {\bf 72}, 035018 (2005)
  doi:10.1103/PhysRevD.72.035018
  [hep-ph/0411263].



\bibitem{CarcamoHernandez:2005ka} 
  A.~E.~Carcamo Hernandez, R.~Martinez and F.~Ochoa,
  Phys.\ Rev.\ D {\bf 73}, 035007 (2006)
  doi:10.1103/PhysRevD.73.035007
  [hep-ph/0510421].



\bibitem{Okada:2016whh} 
  H.~Okada, N.~Okada, Y.~Orikasa and K.~Yagyu,
  Phys.\ Rev.\ D {\bf 94}, no. 1, 015002 (2016)
  doi:10.1103/PhysRevD.94.015002
  [arXiv:1604.01948 [hep-ph]].



\bibitem{Salazar:2015gxa} 
  C.~Salazar, R.~H.~Benavides, W.~A.~Ponce and E.~Rojas,
  JHEP {\bf 1507}, 096 (2015)
  doi:10.1007/JHEP07(2015)096
  [arXiv:1503.03519 [hep-ph]].



\bibitem{Martinez:2008jj} 
  R.~Martinez and F.~Ochoa,
  Phys.\ Rev.\ D {\bf 77}, 065012 (2008)
  doi:10.1103/PhysRevD.77.065012
  [arXiv:0802.0309 [hep-ph]].



\bibitem{Buras:2013dea} 
  A.~J.~Buras, F.~De Fazio and J.~Girrbach,
  JHEP {\bf 1402}, 112 (2014)
  doi:10.1007/JHEP02(2014)112
  [arXiv:1311.6729 [hep-ph]].



\bibitem{Buras:2014yna} 
  A.~J.~Buras, F.~De Fazio and J.~Girrbach-Noe,
  JHEP {\bf 1408}, 039 (2014)
  doi:10.1007/JHEP08(2014)039
  [arXiv:1405.3850 [hep-ph]].



\bibitem{Buras:2012dp} 
  A.~J.~Buras, F.~De Fazio, J.~Girrbach and M.~V.~Carlucci,
  JHEP {\bf 1302}, 023 (2013)
  doi:10.1007/JHEP02(2013)023
  [arXiv:1211.1237 [hep-ph]].



\bibitem{Ponton:2012bi} 
  E.~Ponton,
  arXiv:1207.3827 [hep-ph].



\bibitem{Quiros:2013yaa} 
  M.~Quiros,
  Mod.\ Phys.\ Lett.\ A {\bf 30}, no. 15, 1540012 (2015)
  doi:10.1142/S021773231540012X
  [arXiv:1311.2824 [hep-ph]].



\bibitem{Montero:2000ng} 
  J.~C.~Montero, V.~Pleitez and M.~C.~Rodriguez,
  Phys.\ Rev.\ D {\bf 65}, 035006 (2002)
  doi:10.1103/PhysRevD.65.035006
  [hep-ph/0012178].



\bibitem{CapdequiPeyranere:2001dd} 
  M.~Capdequi-Peyranere and M.~C.~Rodriguez,
  Phys.\ Rev.\ D {\bf 65}, 035001 (2002)
  doi:10.1103/PhysRevD.65.035001
  [hep-ph/0103013].



\bibitem{Diaz:2002dx} 
  R.~A.~Diaz, R.~Martinez and J.~A.~Rodriguez,
  Phys.\ Lett.\ B {\bf 552}, 287 (2003)
  doi:10.1016/S0370-2693(02)03155-6
  [hep-ph/0208176].



\bibitem{Rodriguez:2005jt} 
  M.~C.~Rodriguez,
  Int.\ J.\ Mod.\ Phys.\ A {\bf 21}, 4303 (2006)
  doi:10.1142/S0217751X06031600
  [hep-ph/0510333].



\bibitem{Sen:2005iw} 
  S.~Sen and A.~Dixit,
  hep-ph/0510393.



\bibitem{Sen:2005hj} 
  S.~Sen and A.~Dixit,
  hep-ph/0503078.



\bibitem{Dong:2006vk} 
  P.~V.~Dong, D.~T.~Huong, M.~C.~Rodriguez and H.~N.~Long,
  Eur.\ Phys.\ J.\ C {\bf 48}, 229 (2006)
  doi:10.1140/epjc/s10052-006-0010-z
  [hep-ph/0604028].



\bibitem{Dong:2007qc} 
  P.~V.~Dong, D.~T.~Huong, M.~C.~Rodriguez and H.~N.~Long,
  Nucl.\ Phys.\ B {\bf 772}, 150 (2007)
  doi:10.1016/j.nuclphysb.2007.03.003
  [hep-ph/0701137].



\bibitem{Huong:2008ww} 
  D.~T.~Huong and H.~N.~Long,
  JHEP {\bf 0807}, 049 (2008)
  doi:10.1088/1126-6708/2008/07/049
  [arXiv:0804.3875 [hep-ph]].



\bibitem{Huong:2008ia} 
  D.~T.~Huong and H.~N.~Long,
  Phys.\ Atom.\ Nucl.\  {\bf 73}, 791 (2010)
  doi:10.1134/S1063778810050078
  [arXiv:0807.2346 [hep-ph]].



\bibitem{Huong:2010km} 
  D.~T.~Huong and H.~N.~Long,
  J.\ Phys.\ G {\bf 38}, 015202 (2011)
  doi:10.1088/0954-3899/38/1/015202
  [arXiv:1004.1246 [hep-ph]].



\bibitem{Giang:2012vs} 
  P.~T.~Giang, L.~T.~Hue, D.~T.~Huong and H.~N.~Long,
  Nucl.\ Phys.\ B {\bf 864}, 85 (2012)
  doi:10.1016/j.nuclphysb.2012.06.008
  [arXiv:1204.2902 [hep-ph]].



\bibitem{Hue:2013uw} 
  L.~T.~Hue, D.~T.~Huong and H.~N.~Long,
  Nucl.\ Phys.\ B {\bf 873}, 207 (2013)
  doi:10.1016/j.nuclphysb.2013.04.014
  [arXiv:1301.4652 [hep-ph]].



\bibitem{Ferreira:2013nla} 
  J.~G.~Ferreira, C.~A.~de S.Pires, P.~S.~Rodrigues da Silva and A.~Sampieri,
  Phys.\ Rev.\ D {\bf 88}, no. 10, 105013 (2013)
  doi:10.1103/PhysRevD.88.105013
  [arXiv:1308.0575 [hep-ph]].



\bibitem{Binh:2013axa} 
  D.~T.~Binh, L.~T.~Hue, D.~T.~Huong and H.~N.~Long,
  Eur.\ Phys.\ J.\ C {\bf 74}, no. 5, 2851 (2014)
  doi:10.1140/epjc/s10052-014-2851-1
  [arXiv:1308.3085 [hep-ph]].



\bibitem{Ferreira:2017mvp} 
  J.~G.~Ferreira, C.~A.~de S. Pires, P.~S.~Rodrigues da Silva and C.~Siqueira,
  Eur.\ Phys.\ J.\ C {\bf 78}, no. 3, 225 (2018)
  doi:10.1140/epjc/s10052-018-5705-4
  [arXiv:1706.04904 [hep-ph]].



\bibitem{Bora:2012tx} 
  K.~Bora,
  Horizon {\bf 2} (2013)
  [arXiv:1206.5909 [hep-ph]].



\bibitem{Xing:2007fb} 
  Z.~z.~Xing, H.~Zhang and S.~Zhou,
  Phys.\ Rev.\ D {\bf 77}, 113016 (2008)
  doi:10.1103/PhysRevD.77.113016
  [arXiv:0712.1419 [hep-ph]].



\bibitem{Olive:2016xmw} 
  C.~Patrignani {\it et al.} [Particle Data Group],
  Chin.\ Phys.\ C {\bf 40}, no. 10, 100001 (2016).
  doi:10.1088/1674-1137/40/10/100001



\bibitem{Das:2012ze} 
  A.~Das and N.~Okada,
  Phys.\ Rev.\ D {\bf 88}, 113001 (2013)
  doi:10.1103/PhysRevD.88.113001
  [arXiv:1207.3734 [hep-ph]].



\bibitem{Das:2016hof} 
  A.~Das, P.~Konar and S.~Majhi,
  JHEP {\bf 1606}, 019 (2016)
  doi:10.1007/JHEP06(2016)019
  [arXiv:1604.00608 [hep-ph]].

\end{thebibliography}
\end{document}